\documentclass{article}
\usepackage{graphics,epsfig,amsfonts,amsmath}
\usepackage{placeins}
\usepackage{subfigure}
\begin{document}
\newtheorem{theorem}{Theorem}[section]
\newtheorem{lemma}[theorem]{Lemma}
\newtheorem{proposition}[theorem]{Proposition}
\newtheorem{corollary}[theorem]{Corollary}
\newenvironment{proof}[1][Proof]{\begin{trivlist}
\item[\hskip \labelsep {\bfseries #1}]}{\end{trivlist}}
\newenvironment{definition}[1][Definition]{\begin{trivlist}
\item[\hskip \labelsep {\bfseries #1}]}{\end{trivlist}}
\newenvironment{example}[1][Example]{\begin{trivlist}
\item[\hskip \labelsep {\bfseries #1}]}{\end{trivlist}}
\newenvironment{remark}[1][Remark]{\begin{trivlist}
\item[\hskip \labelsep {\bfseries #1}]}{\end{trivlist}}
\newcommand{\qed}{\nobreak \ifvmode \relax \else
\ifdim\lastskip<1.5em \hskip-\lastskip
\hskip1.5em plus0em minus0.5em \fi \nobreak
\vrule height0.75em width0.5em depth0.25em\fi}
\newcommand{\N}{\mathbb{N}}
\newcommand{\R}{\mathbb{R}}
\newcommand{\Q}{\mathbb{Q}}
\def\beq{\begin{equation}}
\def\eeq{\end{equation}}
\title{Spatiotemporal sine-Wiener Bounded Noise and its effect on Ginzburg-Landau model}
\author{Sebastiano de Franciscis$^{1}$ and Alberto d'Onofrio$^{1}$ (corresponding author)
} % IMPORTANT: leave this curly bracket as the first character of this line.
\date{}
% Date - leave this blank.
\maketitle
%
% Institutions
% Here fill in your institute name(s) and address(es)
% The number in $^...$ indicates the author number. For example
{\center $^1$ European Institute of Oncology, Via Ripamonti 435, Milano, Italy,
I-20141\\\textit{Email address:}
alberto.donofrio@ieo.eu\\[3mm]}
\begin{abstract}
In this work, we introduce a kind of spatiotemporal bounded noise derived by the sine-Wiener noise and by the spatially colored unbounded noise introduced by Garc\'ia-Ojalvo, Sancho and Ram\'irez-Piscina (GSR noise). We characterize the behavior of the distribution of this novel noise by showing its dependence on both the temporal and the spatial autocorrelation strengths. In particular, we show that  the distribution experiences a stochastic transition from bimodality to trimodality. 

Then, we employ the noise here defined to study phase transitions on Ginzburg-Landau model. Various phenomena are evidenced by means of numerical simulations, among which re-entrant transitions, as well as differences in the response of the system to GSR noise additive perturbations. 

Finally, we compare the statistical behaviors induced by the sine-Wiener noise with those caused by 'equivalent' GSR noises.
\end{abstract}
\section{Introduction}
In zero-dimensional nonlinear systems noise may induce a wide spectrum of important phenomena such as stochastic resonance \cite{gammaitoni},
coherence-resonance \cite{lucafrancesco} and noise-induced transitions \cite{hl,wiolindenberg,lucafrancesco}. Noise-induced transitions (also called phenomenological stochastic bifurcations) consists in qualitative changes of the statistical properties of a stochastic system, characterized by transitions from unimodality to bimodality of the stationary probability densities of state variables, and similar phenomena. Note that noise-induced-transitions are well-distinct from phase transitions that need spatially extended systems \cite{wiolindenberg}. Genuine noise-induced phase transitions have been, instead and not surprisingly, found in many spatiotemporal dynamical systems \cite{Ibanhes,GObook,sagues}.

Many studies in the field of noise-induced phenomena in both zero-dimensional and in spatially extended systems were, respectively, based on temporal \cite{hl} or spatiotemporal white noises \cite{sagues, Wang1,Wang2,Wang3}. This important model of noise is, however, mainly appropriate when modeling internal "hidden" degrees of freedom, of microscopic nature. On the contrary, extrinsic fluctuations (i.e. originating externally to the system in study) may exhibit both temporal and spatial structures \cite{GObook,sanchoPhysD}, which may induce new effects. For example, it was shown that zero dimensional systems perturbed by colored noises exhibit correlation-dependent properties that are missing in case of null autocorrelation time, such as the emergence of stochastic resonance also for linear systems, and re-entrance phenomena, i.e. transitions from monostability to bistability and back to monostability \cite{wiolindenberg,hanggi,lucafrancesco}. Even more striking effects are observed in spatially extended systems that are perturbed by spatially white but temporally colored noises. These phenomena are induced by a complex interplay between noise intensity, spatial coupling and autocorrelation time \cite{wiolindenberg}.

Garc\'ia-Ojalvo, Sancho and Ram\'irez-Piscina introduced in \cite{GO92} the spatial version of the Ornstein-Uhlenbeck noise, which we shall call GSR noise, characterized by both a temporal scale $\tau$ and a spatial scale $\lambda$ \cite{lam}.  The Ginzburg-Landau field model - one of the best-studied amplitude equation representing 'universal' nonlinear mechanisms - additively perturbed by the GSR noise was investigated in \cite{GO94,GObook}, where
 it was shown the existence of a non-equilibrium phase transition controlled by both the correlation time and the correlation length \cite{GO94,GObook}.

The above-summarized body of research is essentially based on the use of Gaussian Noises (GNs), which is the best approximation of reality in many cases. 
However, an increasing number of experimental data shows that many real-life stochastic processes does not follow white or colored Gaussian laws, but other probability densities (such as “fat-tail” power-laws \cite{New05}). More recently, theoretical research focused on another important class of non-Gaussian stochastic processes: the bounded noises. 
Probably the most studied  bounded noise is the Dichotomous Markov Noise (DMN)\cite{lucafrancesco}. In the last twenty years, other classes of bounded noises were defined and intensively studied in statistical physics \cite{wioII,CaiLin,bobryk} and in engineering \cite{dimentberg}, and - to a lesser degree - in mathematics \cite{Homburg} and quantitative biology \cite{pre,dongan}.

The rise of scientific interest on bounded noises is motivated by the fact that in many applications both GNs and “fat-tailed” non-Gaussian stochastic processes are an inadequate mathematical model of the physical world because of their infinite domain. This should preclude their use to model stochastic fluctuations affecting parameters of dynamical systems, which must be bounded by physical constraints \cite{wioII,bobryk,pre}. Moreover, in biology and elsewhere, some parameters must also be strictly positive. As a consequence, not taking into account the bounded nature of stochastic fluctuations may lead to unrealistic inferences. For instance, when the onset of noise-induced transitions depends on exceeding a threshold by the variance of a GN, this often means making negative or excessively large a parameter \cite{wioII,bobryk,dongan,pre}. To give an example taken from medicine, a GN-based modeling of the unavoidable fluctuations affecting the pharmacokinetics of an antitumor drug delivered by means of continuous infusion leads to a paradox. Indeed, the probability that the drug increases the number of tumor cells may become nonzero, which is absurd \cite{dongan,pre}. Thus, in order to avoid these problems, the stochastic models should in these cases be built on bounded noises. 

In order to generate a temporal bounded noise, two basic recipes have been adopted so far. The first consists in generating the noise by means of an appropriate stochastic differential equation \cite{wioII,CaiLin}, whereas the second one consists in applying a bounded function to a standard Wiener process. In the purely temporal setting, two relevant examples of noises obtained by implementing the fist recipe are the Tsallis-Borland  \cite{wioII} and the Cai-Lin \cite{CaiLin} noises, whereas an example generated by following the second recipe is the zero-dimensional sine-Wiener noise \cite{bobryk}.

Recently, in \cite{deFradOnpre} we introduced and numerically studied two spatiotemporal extensions of the above-mentioned Tsallis-Borland and Cai-Lin noises. In that work we applied - as an additive perturbation - these bounded noise to a Ginzburg-Landau (GL) model and stressed out the dependence of the phase transitions phenomena on both spatial and temporal correlation strength, as well as on the specific model of noise that has been adopted.

Our aim here is threefold. First, by adopting the 'second recipe' we want to define and numerically investigate a new simple spatiotemporal bounded noise, which extends both the zero-dimensional sine-Wiener bounded noise \cite{bobryk}, and the spatiotemporal unbounded GSR noises \cite{GO92,GObook}.

Second, we want to assess the effects of such bounded stochastic forces (i.e. of additive bounded noises) and of initial conditions on the statistical properties of the spatiotemporal dynamics of the Ginzburg-Landau (GL) equation. 

Third we explore the similar and different features of the spatiotemporal sine-Wiener noise perturbation with respect to those of the  Cai-Lin and Tsallis-Borland spatiotemporal bounded noises studied in \cite{deFradOnpre}.

Phase transitions induced in GL model by additive and multiplicative unbounded noises were extensively studied in last twenty years \cite{GObook,GO92pla,GO92,gpsv,various,ms,jstat,ss,lucafrancesco,ouch}. It follows that we shall mainly focus on the effects more strictly related to the boundeness of the noise in study. In particular, we will compare the response of GL system to SW noise with the one to GSR noise.

\section{Spatiotemporal colored unbounded noise}
Let us consider the well-known zero-dimensional Ornstein-Uhlenbeck stochastic differential equation:
\begin{equation}\label{oue}
\xi^{\prime} (t)= -\frac{1}{\tau}\xi(t) + \frac{\sqrt {2 D} }{\tau}\eta(t),
\end{equation}
where $\tau$ is the typical temporal correlation, $\sqrt {2 D}$  is the noise strength and $\eta(t)$ is a Gaussian white noise of unitary intensity:
\begin{equation}
\langle \eta(t)\eta(t_1)\rangle= \delta(t-t_1).
\end{equation}
It is well-known that solution of eq. (\ref{oue}) is a gaussian colored stochastic process with autocorrelation:
\begin{equation}
\langle \xi(t)\xi(t_1)\rangle \propto \ exp\left(-\frac{|t-t_1|}{\tau}\right).
\end{equation}
In \cite{GO92} eq. (\ref{oue}) 
was generalized in a spatially extended setting by including in it the most known and simple spatial coupling, the Laplace operator, yielding the following partial differential Langevin equation
\begin{equation}\label {gener}
\partial_t \xi (x,t)= \frac{\lambda^2}{2 \tau}\nabla^2 \xi(x,t) -\frac{1}{\tau}\xi(x,t) + \frac{\sqrt {2 D} }{\tau}\eta(x,t),
\end{equation}
where $\lambda>0$ is the spatial correlation strength \cite{GO92} of $\xi (x,t)$.\\
As usual in non-equilibrium statistical physics, we shall investigate the lattice version of (\ref{gener}):
\begin{equation}\label {generlattice}
\xi_p^{\prime} (t)= \frac{\lambda^2}{2 \tau}\nabla_L^2 \xi_p(t) -\frac{1}{\tau}\xi_p(t) + \frac{\sqrt {2 D} }{\tau}\eta_p(t),
\end{equation}
where $ p = h \ (i,j)$ is a point on a $N*N$ lattice with steps equal to $h$. The symbol $\nabla_L^2$ denotes the discrete version of the Laplace's operator:
\begin{equation}\label {lapllatt}
\nabla_L^2 \xi_p (t)= \frac{1}{h^2}\sum_{i \in ne(p)}(\phi_i-\phi_p),
\end{equation}
where $ne(p)$ is the set of the neighbors of the lattice point $p$. The Weiss mean field method \cite{parrondo} applied to eq. (\ref{generlattice}) yields that for $N>>1$ the one-site distribution of the GSR noise is:$ P_{GSR}(\xi) = C exp(-\xi^2/(2 \sigma_{GSR}^2) )$, where
\begin{equation}\label{GSRsigma}
\sigma_{GSR}^2 = \frac{D}{\tau_c(1+2 \lambda^2)}
\end{equation}
\section{The sine-Wiener spatiotemporal bounded noise: definition and properties}\label{SWnoise}
The sine-Wiener noise is obtained by applying the bounded function $h(u) = B \sin(\sqrt{2/\tau}u)$ to a random walk $W(t)$ defined as $W'= \eta(t)$, where $\eta(t)$ is a white noise of unitary intensity, yielding:
\begin{equation}
\zeta(t)= B \sin\left( \sqrt{ \frac{2}{\tau}}W(t) \right).
\end{equation}
The stationary probability density of $\zeta(t)$ is given by
\begin{equation}
P_eq(\zeta) = \frac{1}{\pi \sqrt{B^2 -\zeta^2}}, 
\end{equation}
thus: $P_eq(\pm B)=+\infty$. Thanks to this property, one may consider the sine-Wiener noise as a realistic extension of the Markov dichotomous noise, whose stationary density is $P_eq(\zeta) = (1/2)\delta(\zeta-|B|)$. 

Here, as a natural spatial extension of the sine-Wiener noise, we define the following spatiotemporal noise:
\begin{equation}\label{XXX}
\zeta(x,t)= B \sin\left( 2 \pi \xi(x,t) \right),
\end{equation}
where $\xi(x,t)$ is the spatiotemporal correlated noise defined by (\ref{gener}). 

If the number of lattice sites is sufficiently large, we may study the global behavior of the spatiotemporal noise by means of the equilibrium heuristic probability density of the noise lattice variables $\zeta_p$, $P_{eq}(\zeta)$.

\begin{figure}
\begin{center}
\subfigure[]
{
\label{A}
\includegraphics[width=0.48\textwidth]{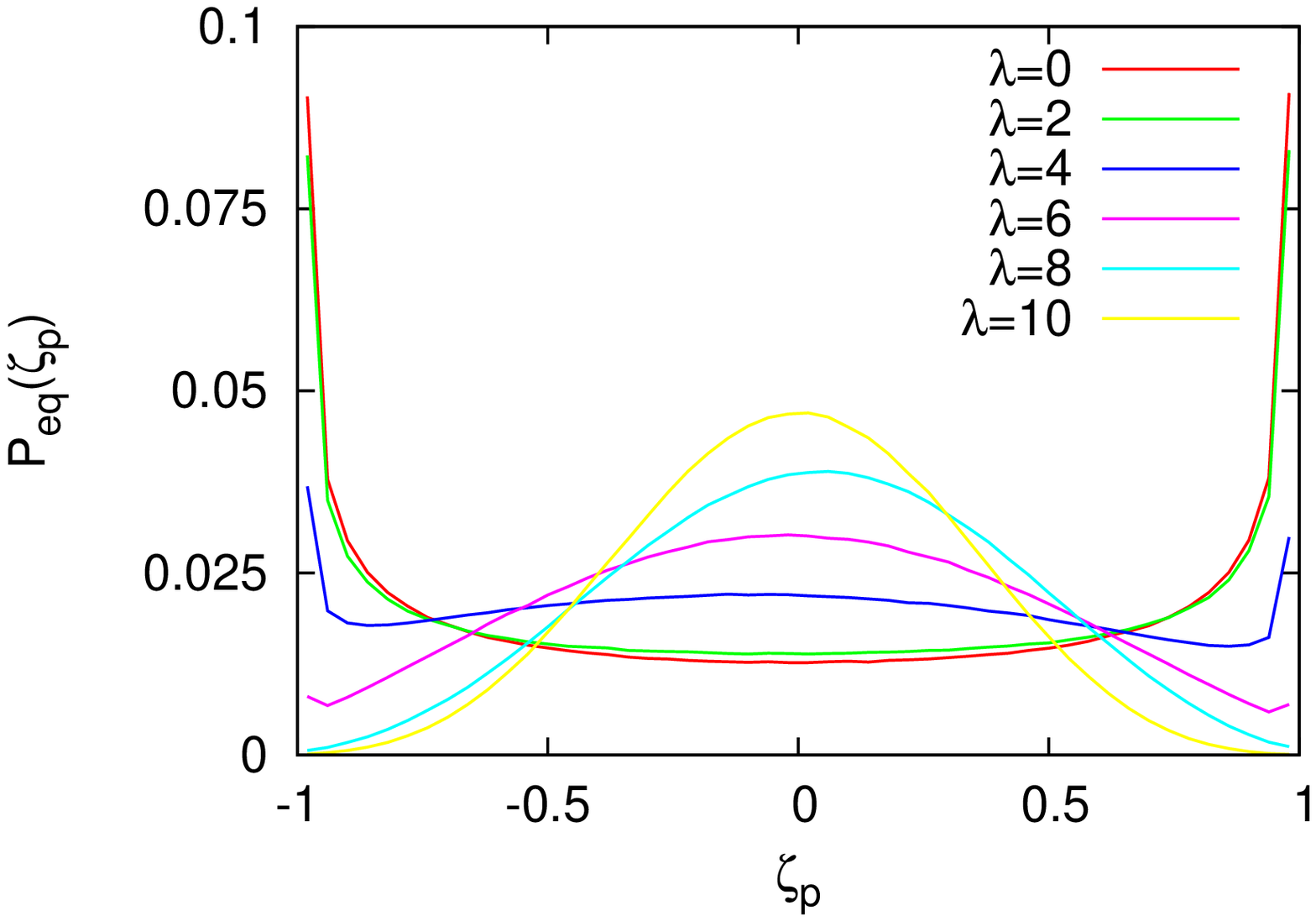}
}
\subfigure[]
{
\label{B}
\includegraphics[width=0.48\textwidth]{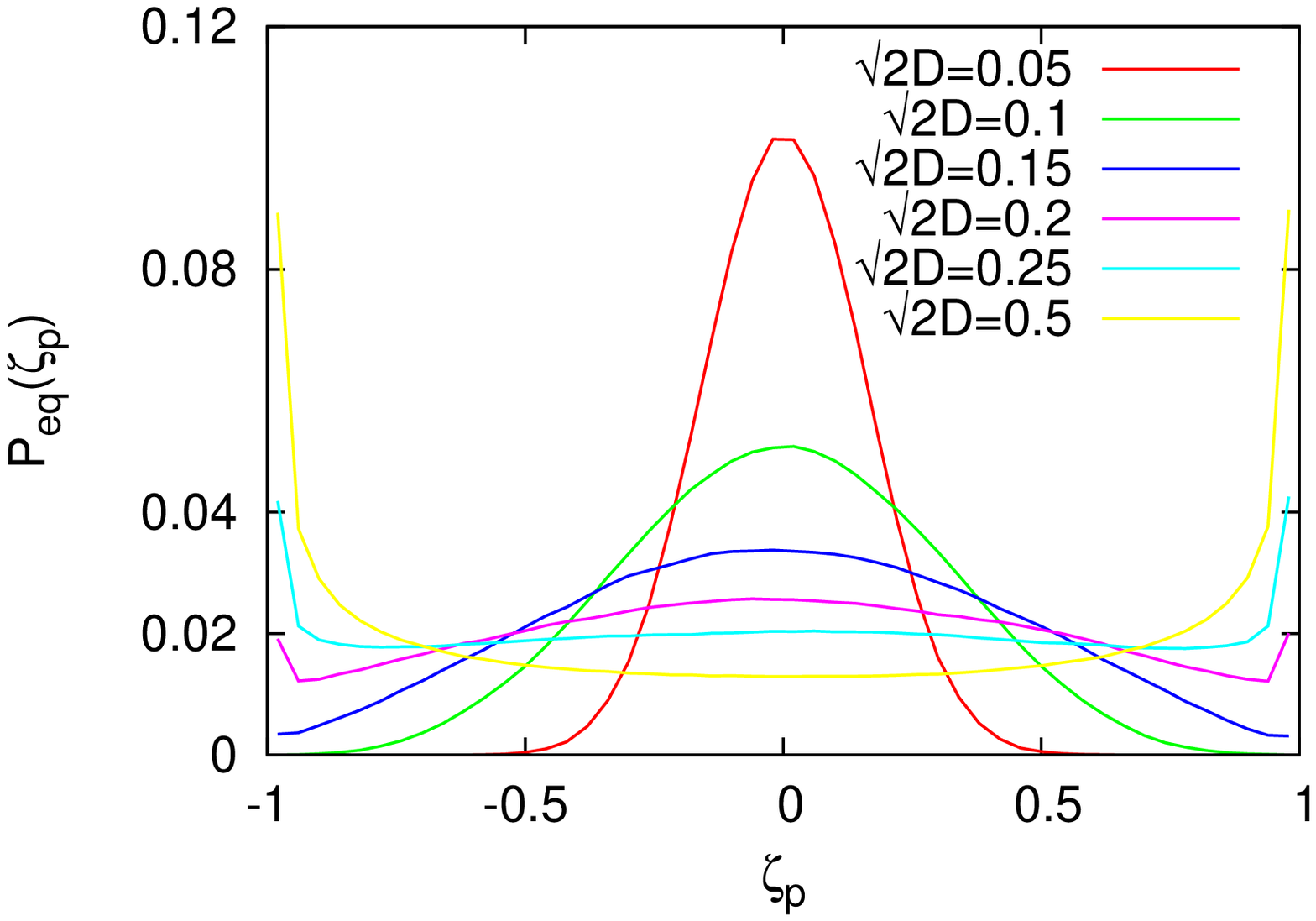}
}
\end{center}
\caption{Equilibrium distribution $P_{eq}(\zeta_p)$  of the sine-Wiener bounded spatiotemporal noise, on a $40\times40$ lattice system with $B=1$. Panel \subref{A}: stochastic bifurcation induced by varying the parameter $\lambda$, with $\tau_c=2$ and $\sqrt{2D} =1$. Panel \subref{B}: stochastic bifurcation induced by varying the parameter $\sqrt{2D}$, with $\tau_c=2$ and $\lambda=0$.}
\label{fig_P_eq}
\end{figure}

We observed that when varying the spatial coupling parameter $\lambda$ of the underlying GSR noise, the distribution of $\zeta(x,t)$ exhibits at $\lambda = \lambda^* \approx 4 $ a stochastic bifurcation (see figure \ref{fig_P_eq}.a): for $0 \le \lambda < \lambda^*$ the distribution is bimodal, whereas for $\lambda>\lambda^*$ the distribution is trimodal, since an additional mode at $\zeta = 0$ appears. Similar bifurcations are observed if varying $D$ (see figure \ref{fig_P_eq}.b) or $\tau$ (although, in this case, the bifurcation value is very large). 

These behaviors may be heuristically explained  by the one-site distribution of the underlying GSR noise $\xi$. Indeed, defining the 'span' of the GSR noise as 
\begin{equation}\label{sp}
S = 2 \sigma_{GSR} = 2 \sqrt{\frac{D}{\tau_c(1+2 \lambda^2)}}
\end{equation}
yields that $S$ increases with $D$, and decreases both with $\tau_c$ and $\lambda$. Thus both the above-mentioned numerically observed phenomena may be explained. 

To start, note that for small $\lambda$ it is $S \approx \sqrt{2 D/\tau_C}$. Thus if $S$ is sufficiently large, the horn-shaped distribution is observed, due to the large span of the argument of the sinus, which remains roughly constant (provided that $\lambda$ is such that $2 \lambda^2<<1$). 

On the contrary, for large $\lambda$, $S$ becomes small, and the argument of the sinus remains prevalently small, whereby causing the onset of a central new mode. 

%%\FloatBarrier
\section{The Ginzburg-Landau equation perturbed by additive sine-Wiener noise}
Let us consider the following bidimensional lattice-based  Ginzburg-Landau equation:
\begin{equation}\label {GLequation}
\partial_t \psi_p= \frac{1}{2}\left(\psi_p-\psi_p^{3}+\nabla_L^2 \psi_p\right) +A_p(t),
\end{equation}
where $A_p(t)$ is a generic bounded or unbounded additive noise. In \cite{GObook} Garc\'ia-Ojalvo \textit{et al.} studied the eq. (\ref{GLequation}) under the assumption that $A_p(t)= \xi_p(t)$, where $\xi_p(t)$ is the GSR noise defined by eq. (\ref{generlattice}). They showed that both spatial and temporal correlation parameters (respectively, $\lambda$ and $\tau$) shift the transition point towards larger values. 

In the following we will illustrate some analytical and numerical results for the case $A_p(t) = \zeta_p(t)$, where $\zeta_p(t)$ is the bounded sine-Wiener noise defined by eq. (\ref{XXX}) and computed at the lattice sites. We stress here that our aim is to provide a solid testbed to the novel type of spatiotemporal bounded noise here defined, and not to evidence some unknown aspects of the GL model, which is one of most important and studied models of statistical mechanics.

In line with \cite{GObook}, phase transitions in GL equation will be characterized by means of the order parameter 'global magnetization', i.e.: 
\begin{equation}
M\equiv\frac{<|\sum_{p}\psi_p|>}{N^{2}},
\end{equation} 
and of its relative fluctuation $\sigma_M$ \cite{GObook}:
\begin{equation}
\sigma_M\equiv \sqrt{\frac{<|\sum_{p}\psi_p|^{2}>-<|\sum_{p}\psi_p|>^{2}}{N^{2}} }.
\end{equation}
Again in line with \cite{GObook}, we define a transition from large to small values of the order parameter as an 'order to disorder' transition. However, by no means we state the equivalences 'disorder = randomness' and 'order = homogeneity'.

All simulations have been performed in a $40\times40$ lattice for a time interval $[0, 250]$, and the temporal averages were computed in the interval $[125,250]$. In all cases, noise initial condition was set to $0$.

We will main focus on the case where the initial state is $\psi(x,0)=1 \mbox{} \forall x$.

\subsection{Some analytical considerations on the role of $B$}
Lattice-based system (\ref {GLequation}) is endowed by an important mathematical property. Indeed, it is a cooperative system \cite{coppel} since:
\begin{equation} \partial_{\psi_k}\psi_p^{\prime}  \ge 0. \end{equation}
This property and the fact that $A_p(t) \ge -B$ implies that: \begin{equation} \psi_p(t)\ge \widetilde{\psi}_p(t), \end{equation}
where
\begin{equation} \partial_t \widetilde{\psi}_p= \frac{1}{2}\left(\widetilde{\psi}_p-\widetilde{\psi}_p^{3}+\nabla_L^2 \widetilde{\psi}_p\right) -B  \end{equation}
with $\widetilde{\psi}_p(0)=\psi_p(0)$.

Now, note that if $0<B<B^* =1/(3 \sqrt{3})$ then the equation
\begin{equation} s-s^3 = 2 B \end{equation}
has three solutions $s^a(B)<0$, $s^b(B)\in (0,1)$ and $s^c(B)\in (0,1)$ such that $s^b(B)<s^c(B)$. For example, for $B=0.19<B^*$ it is: $s^a(0.19)\approx-1.15306$, $s^b(0.19)\approx 0.52331$ and $s^c(0.19)=0.62975$. In particular, if $B<<1$ then it is $s^c(B)\approx 1-B $ and $s^a(B)\approx -1-B$.  It is an easy matter to show that if $\widetilde{\psi}_p(0)>s^b(B)$ then $\widetilde{\psi}_p(t)>s^b(B)$, also implying $\psi_p(t)>s^b(B)$ and of course that $M(t)> s^b(B)$ and $M_s(t)> s^b(B)$. Indeed, suppose that at a given time instant $t_1$ all $\widetilde{\psi}_p(t_1) \ge s^b(B), $ but a point $q$ where $\psi_q(t_1)=s^b(B)$. Thus, it is 
\begin{equation} \partial_t \widetilde{\psi}_q(t_1)= \frac{1}{2}\left(\widetilde{\psi}_q-\widetilde{\psi}_q^{3}+\nabla_L^2 \widetilde{\psi}_q\right) -B = 0 + \frac{1}{2} \nabla_L^2 \widetilde{\psi}_q \ge 0. \end{equation}
Note that the vector $c(B)= s^c(B) (1,\dots,1)$ is a locally stable equilibrium point for the differential system ruling the dynamics of $\widetilde{\psi}_p(t)$. Indeed, $c$ is a minimum of the associated energy. However, the system might be multistable, similarly to the GL model with total coupling in the lattice \cite{rt}. By adopting a Weiss mean field approximation, one can proceed
as in \cite{rt} and infer that the equilibrium is unique for $N>>1$. Namely, defining the auxiliary variable:
\begin{equation} m_p = \sum_{j \in ne(p)} \widetilde{\psi}_j  \end{equation}
the equilibrium equations reads
\begin{equation}\widetilde{\psi}_p^3 + 3 \widetilde{\psi}_p = 4 m_p- 2 B. \end{equation}
Note that we are only interested to the subset $\widetilde{\psi}_p \ge s^b(B)$ that also implies $m_p \ge s^b(B)$. Note now that the equation $s+3 s^3 = x$ for $x>0$ has a unique positive solution $s= k(x)$. Thus
\begin{equation} \widetilde{\psi}_p = k( 4 m_p -  2 B ). \end{equation}
Now, by the following approximation
\begin{equation} m_p \approx \frac{1}{N}\sum_{j=1}^N \widetilde{\psi}_j, \end{equation}
one gets the equation:
\begin{equation} m = k( 4 m - 2 B ),  \end{equation}
which has to be solved under the constraint $m> s^b(B)$. As it is easy to verify, the above equation has only one solution, $m=s^c(B)$.

Any case for $B<<1$ the initial point $\psi_p(0) = 1$ should be such that $\psi_p(t)$ remains in the basin of attraction of $c(B)$, so that for large times $\psi_p(t) \rightarrow s^c(B)$, implying that 
\begin{equation} 
LimInf_{t\rightarrow +\infty}\psi_p(t) \ge s^c(B). 
\end{equation}
 
From the inequality $A_p(t) \le B$, by using similar methods one may infer that for small $B$ it is
\begin{equation} 
LimSup_{t\rightarrow +\infty}\psi_p(t) \le u^c(B). 
\end{equation}
where $u^c(B) > 1$ is the unique positive solution (for $B<B^*$) of the equation 
\begin{equation} u-u^3 = - 2 B. \end{equation}
Note that it is $u^c(B)=-s^a(B)$, due to the anti-symmetry of function $s-s^3$. 
Summing up, we may say that for small $B$ and probably for all $B \in (0,B^*)$ ) it is asymptotically 
\begin{equation} s^c(B) < \psi_p(t) < u^c(B).  \end{equation}

Finally, we numerically solved the system
\begin{equation} \frac{1}{2}\left(\widetilde{\psi}_p-\widetilde{\psi}_p^{3}+\nabla_L^2 \widetilde{\psi}_p\right) -B  = 0\end{equation}
for various values of $B$ in the interval $(0.01,B^*)$ and in all cases we found only one equilibrium with components greater than $s^b(B)$: $\widetilde{\psi}= c(B) = s^c(B)(1,\dots,1) $.  Similarly, when setting $A_p(t)=+B$ in eq. (\ref {GLequation}), we found only one equilibrium value: $u^c(B)(1,\dots,1)$.

\subsection{Phase Transitions}
\begin{figure}
\begin{center}
\subfigure[]
{
\label{A}
\includegraphics[width=0.48\textwidth]{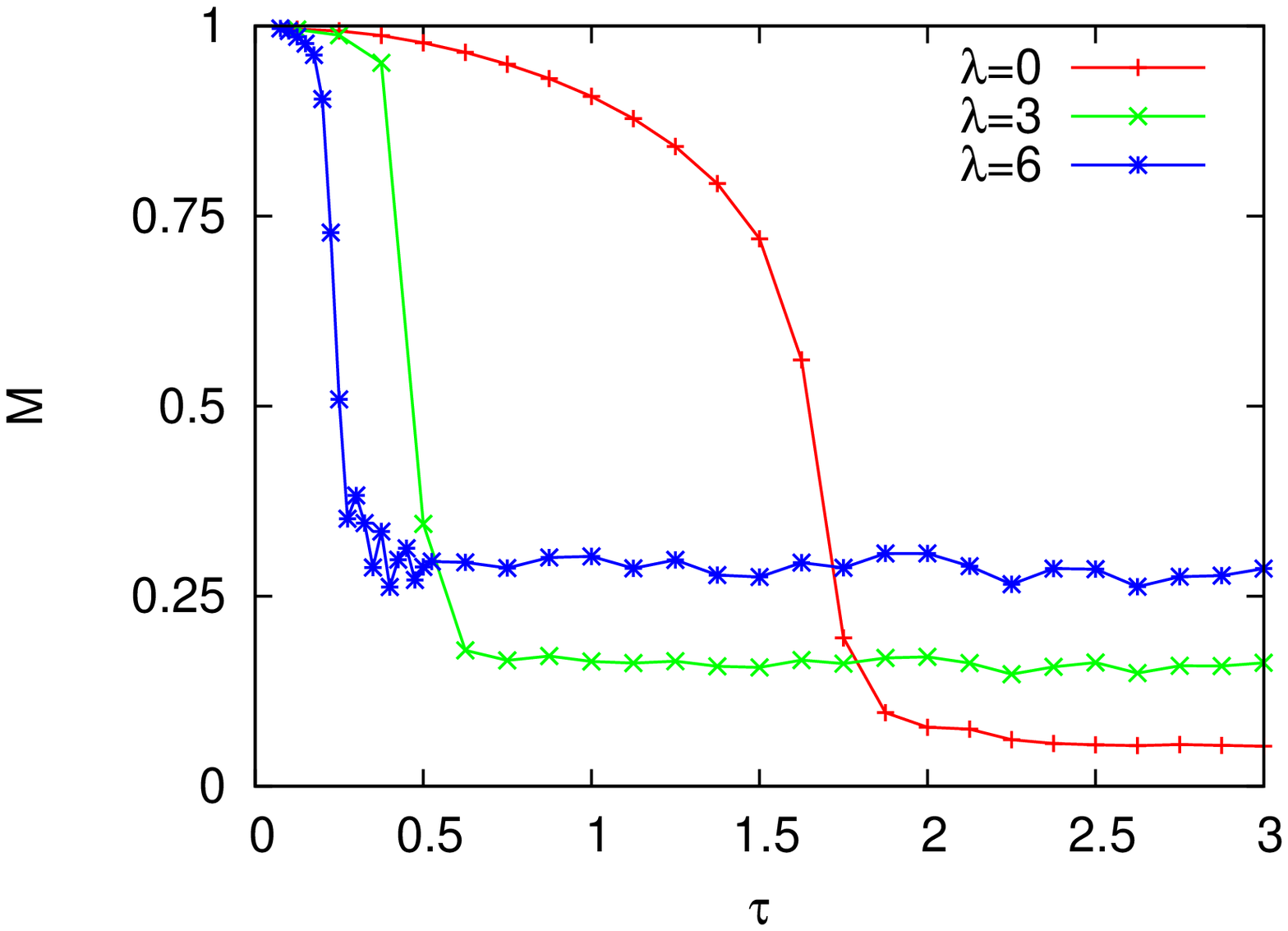}
}
\subfigure[]
{
\label{B}
\includegraphics[width=0.48\textwidth]{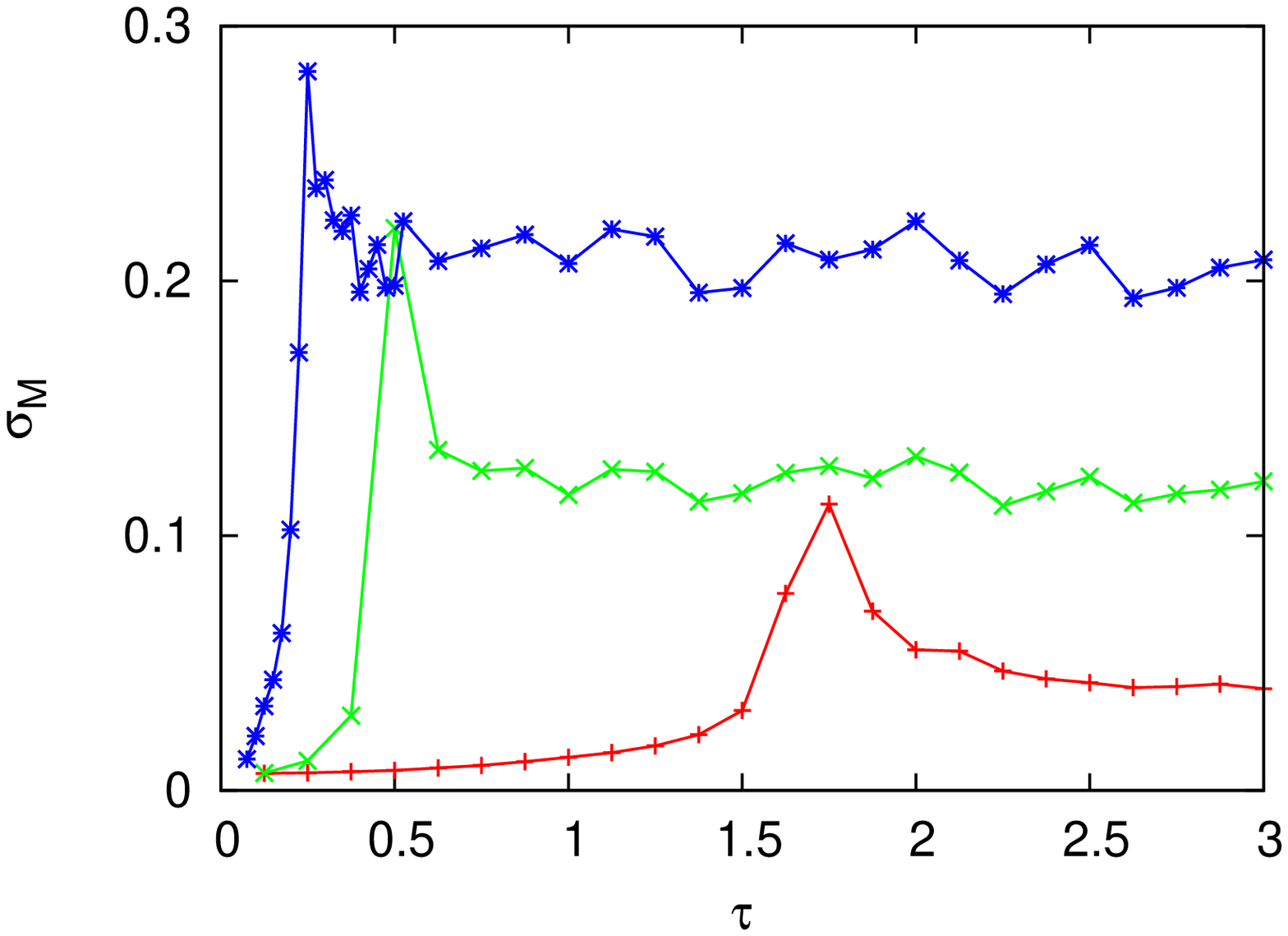}
}
\end{center}
\caption{Effects of autocorrelation parameter $\tau$ on GL model perturbed by additive spatiotemporal sine-Wiener noise. The initial condition is $\psi(x,0)=1$. Panel \subref{A}: global magnetization $M$. Panel \subref{B}: relative fluctuation $\sigma_M$. Other parameters: $B=2.4$ and $\sqrt{2D}=1$.}
\label{fig_GLtau}
\end{figure}

In the curve $M$ vs. $\tau$ a phase transition is observed (see fig. \ref{fig_GLtau}) from large to small values of the order parameter $M$ (a so-called 'order' to 'disorder' phase transition). In absence of spatial autocorrelation, for large $\tau$ it is $M \approx 0$, whereas if one increases $\lambda$ one observe that the lower value of $M$ increases. Moreover, the transition point decreases with increasing $\lambda$.
 
\begin{figure}
\begin{center}
\includegraphics[width=0.5\textwidth]{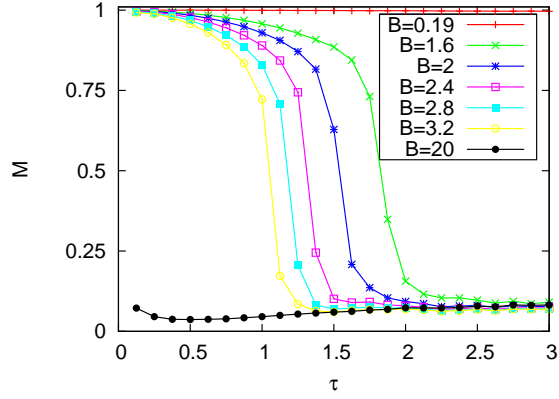}
\end{center}
\caption{Effect of the noise amplitude $B$ on the curve $M$ vs. $\tau$ for GL model perturbed by additive spatiotemporal sine-Wiener noise. Here the initial condition is $\psi(x,0)=1$. Other parameters: $\lambda=1$ and $\sqrt{2D}=1$. }
\label{fig_GLtauB}
\end{figure}

%\FloatBarrier
Figure \ref{fig_GLtauB} shows the influence of the noise amplitude $B$ on curve $M$ vs. $\tau$. We observe that for small $B$, in line with our analytical calculations, no phase transition occurs. For larger $B$, phase transition is observed, and the transition point decreases with increasing noise amplitude.

%\FloatBarrier

Note that, based on the analytical study of the previous subsection, it is excluded that for small values of $B$ a phase transition could be observed for values of $\tau$ that are larger than the ones considered in figure \ref{fig_GLtauB}, also if we change $\lambda$ or $D$. 

Finally, it is interesting to observe that since for large $\tau$ the span $A$ slowly tends to zero, it follows that $M$ will smoothly approach the value $M=1$.

\begin{figure}
\begin{center}
\subfigure[]
{
\label{A}
\includegraphics[width=0.48\textwidth]{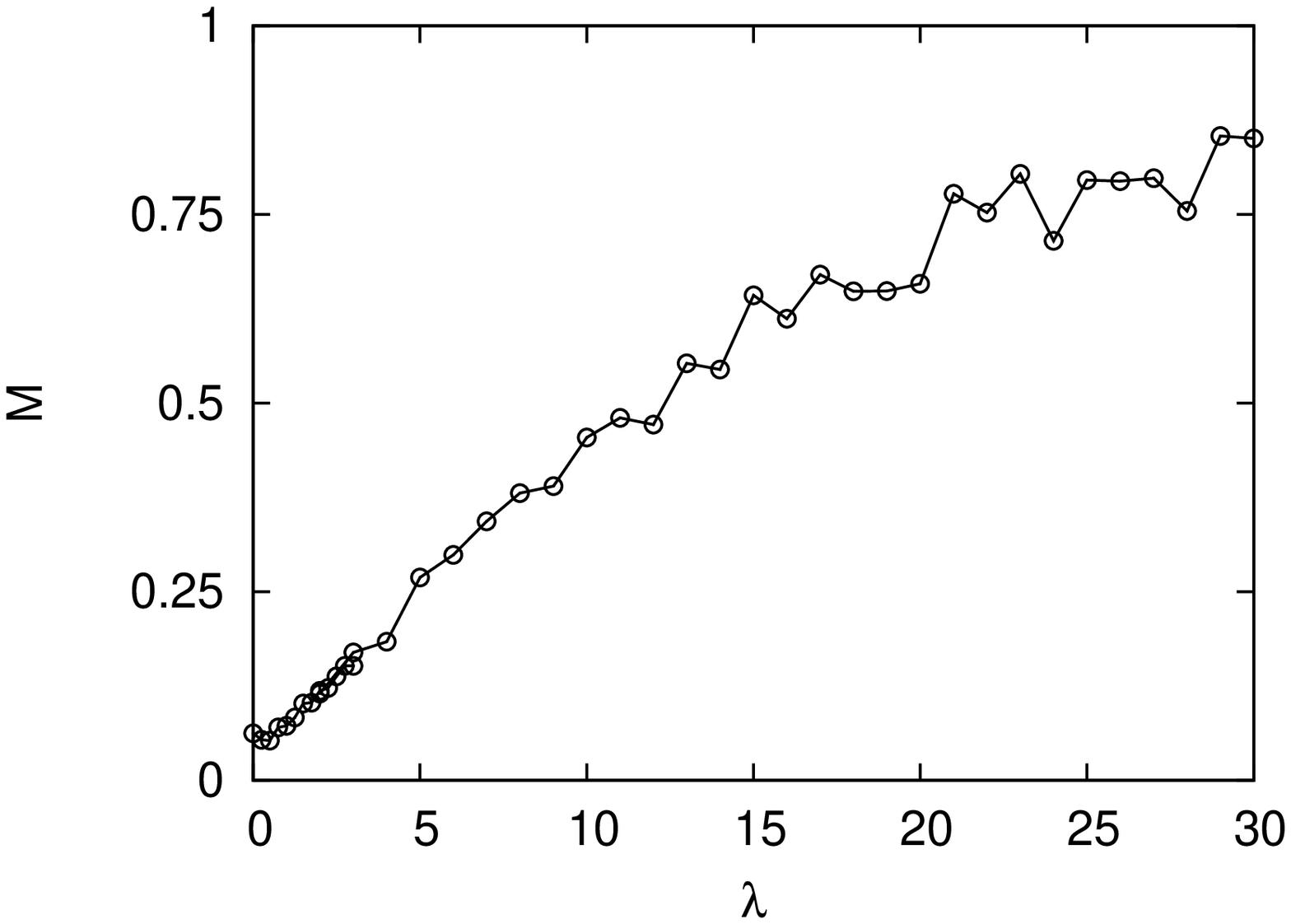}
}
\subfigure[]
{
\label{B}
\includegraphics[width=0.48\textwidth]{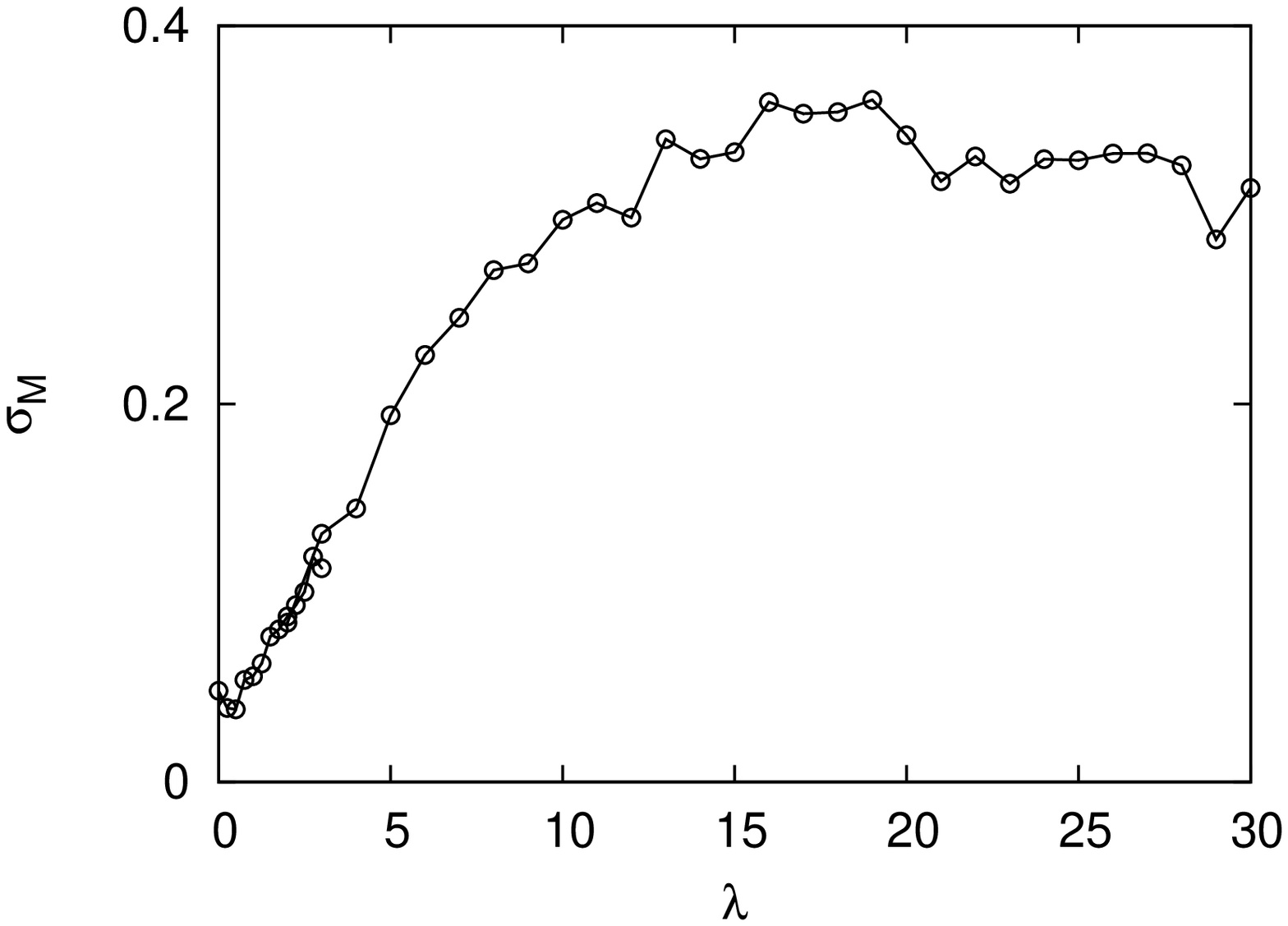}
}
\end{center}
\caption{Effects of spatial autocorrelation coefficient $\lambda$ on GL model perturbed by additive sine-Wiener spatiotemporal noise. Initial condition: $\psi(x,0)=1$. Panel \subref{A}: global magnetization $M$. Panel \subref{B}: relative fluctuation $\sigma_M$. Other parameters are $B=2.6$, $\tau=2$ and $\sqrt{2D}=1$.}
\label{fig_GLlambda}
\end{figure}

Note that an increase of $\lambda$ also causes a decrease of $A$ and in turn a smooth increase of $M$, which can be observed in figure \ref{fig_GLlambda}.(a), where we plot $M$ versus $\lambda$ for $\tau = 2$.  

\begin{figure}
\begin{center}
\includegraphics[width=11.5cm,angle=-90]{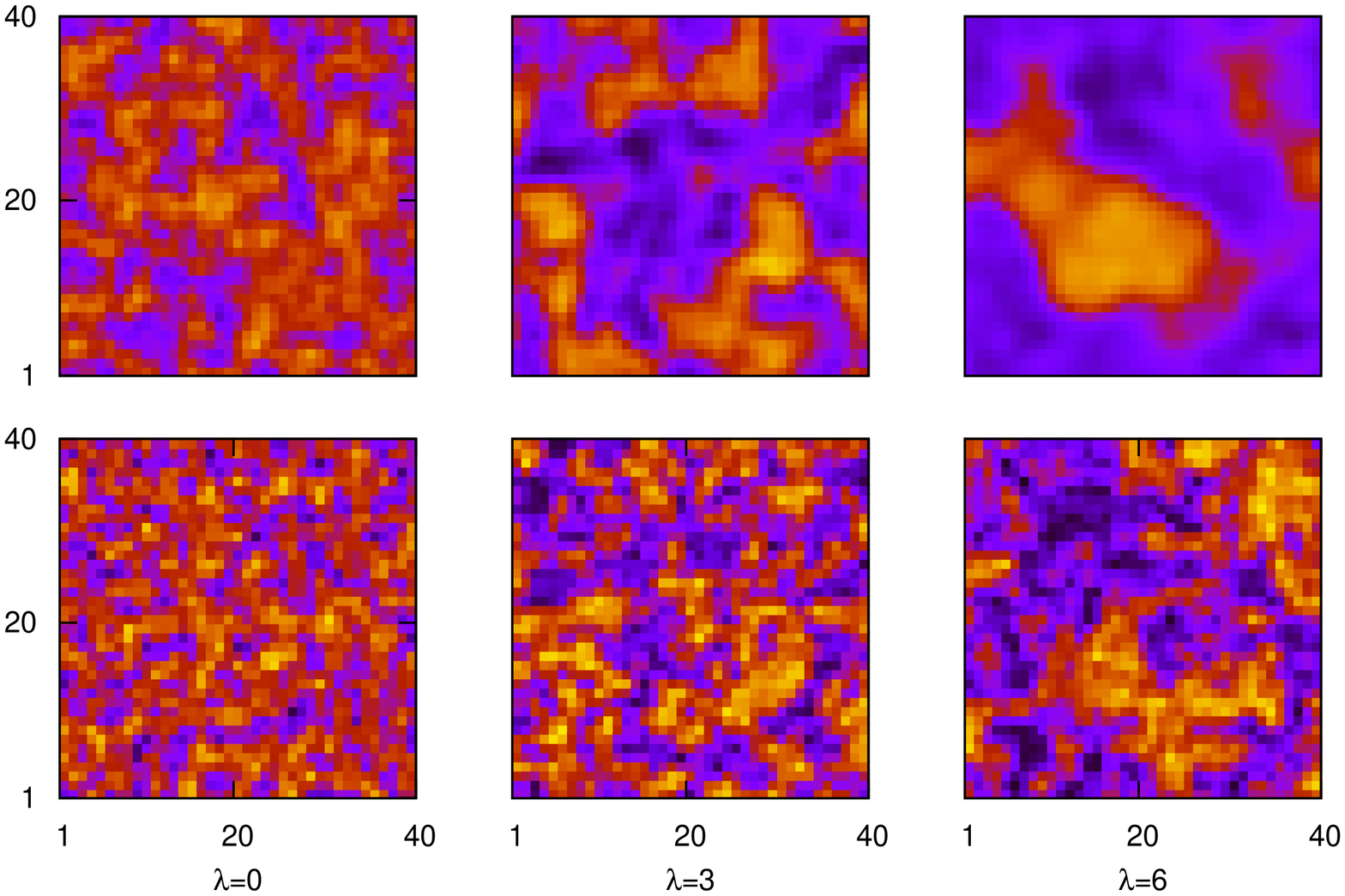}
\end{center}
\caption{Effects of spatial correlation strength $\lambda$ on the field $\psi$ of a $40\times40$ lattice GL system perturbed by additive sine-Wiener noise. The increase of spatial autocorrelation of the field is driven by that of the noise. All the panels refer to $t=250$.  Upper panel: GL field; lower panel: sine-Wiener noise. Other parameters: $B=2.6$, $\sqrt{2D}=0.75$ and $\tau=2$.}
\label{fig_surf}
\end{figure}

In figure \ref{fig_surf}
we show, for three values of $\lambda$, the corresponding heat-map plots of both GL lattice field $\psi$ and of the SW noise $\zeta$. Note that, as one may read in fig \ref{fig_GLlambda}, although the corresponding values of $M$ are not large and one would be tempted to say that the field is 'disordered', the heath-maps instead show large spatially autocorrelated regions, whose size increases with $\lambda$. For example, the 'low' value $M=0.23$ corresponds to the right-upper panel of figure $\ref{fig_surf}$.

%\FloatBarrier

\begin{figure}
\begin{center}
\subfigure[]
{
\label{A}
\includegraphics[width=0.48\textwidth]{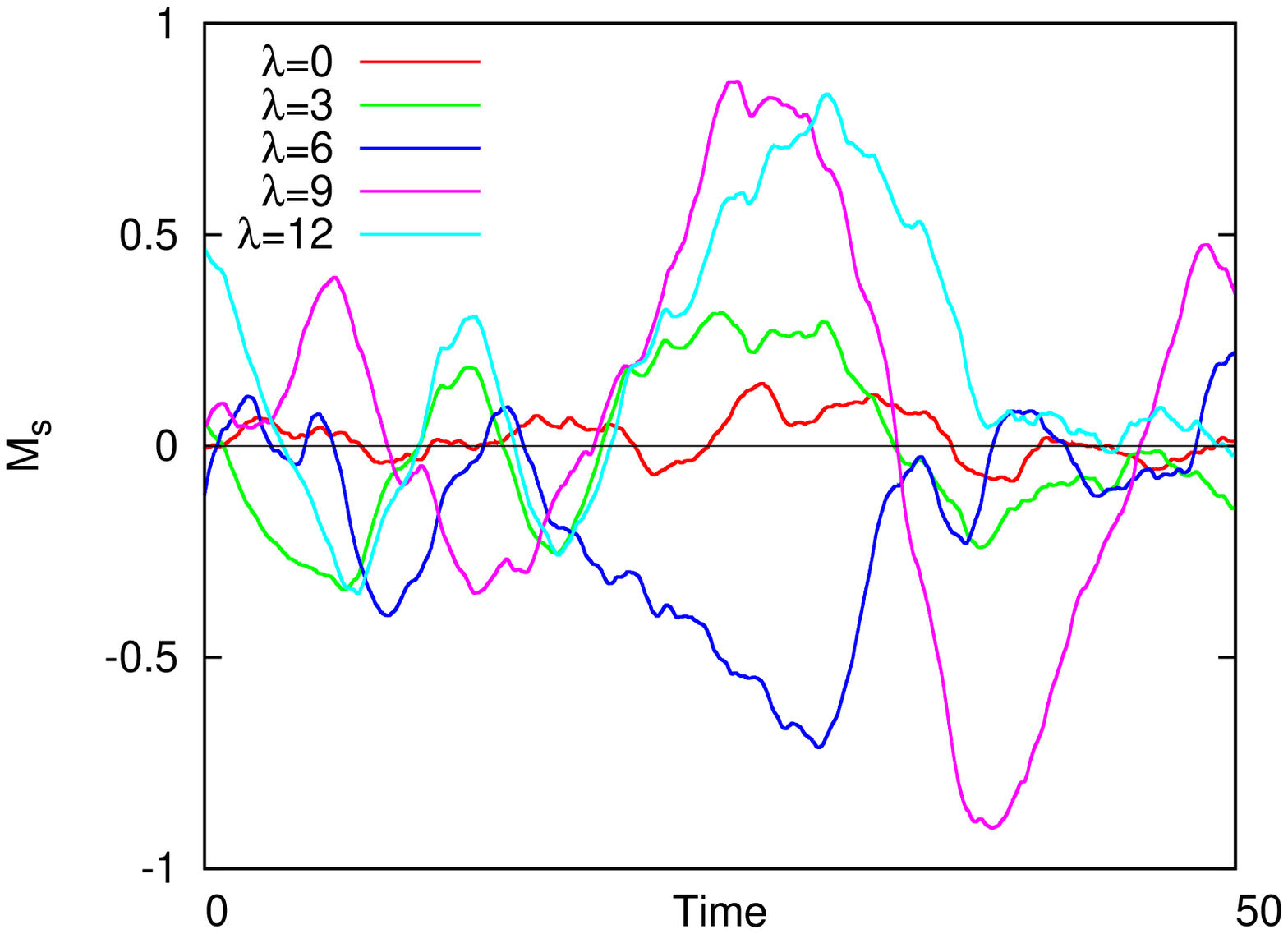}
}
\subfigure[]
{
\label{B}
\includegraphics[width=0.48\textwidth]{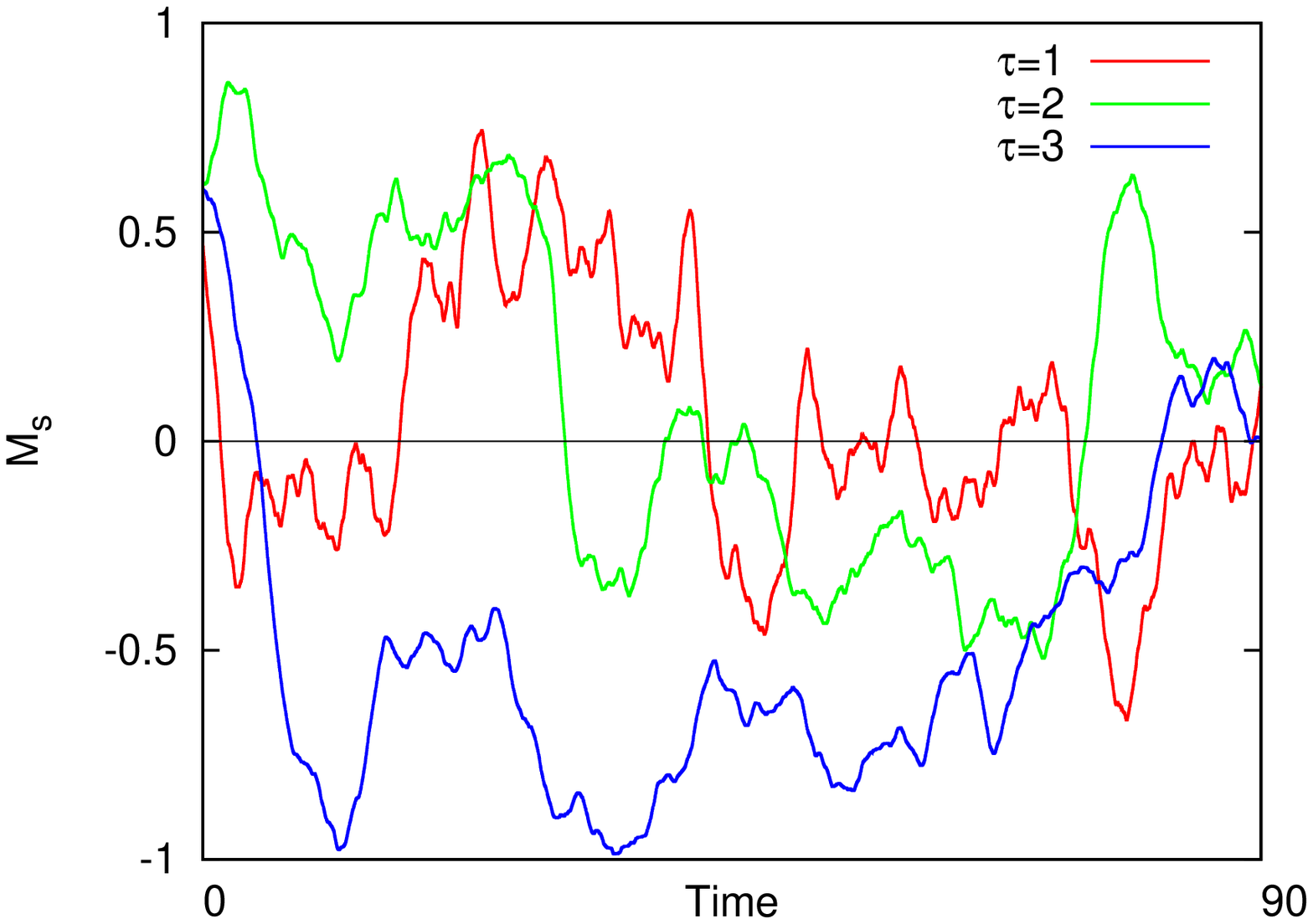}
}
\end{center}
\caption{Effects of spatial and temporal noise correlation parameters in the solutions of GL system, measured by the signed magnetization $M_s$. Panel  \subref{A}: the spatial correlation increases the amplitude of the oscillations between positive and negative signed magnetization (here $\tau = 2.5$). Panel \subref{B}: the parameter $\tau$ decreases the number of switches between negative and positive values $M_s$ (here $\lambda =9$). Other parameters $B=2.4$ and $\sqrt{2D}=1$.}
\label{fig_osc}
\end{figure}

Figure \ref{fig_osc} shows the time series for the signed magnetization defined as 
\begin{equation}
M_s(t)\equiv\frac{\sum_{p}\psi_p(t)}{N^{2}}.
\end{equation}
This figure supports the idea that the above-mentioned clusters are in general non symmetric (i.e. the total positive and negative magnetization is different) and unstable, resulting in an oscillation between positive and negative magnetization, whose amplitude is increasing with $\lambda$.

%\FloatBarrier

\begin{figure}
\begin{center}
\subfigure[]
{
\label{A}
\includegraphics[width=0.48\textwidth]{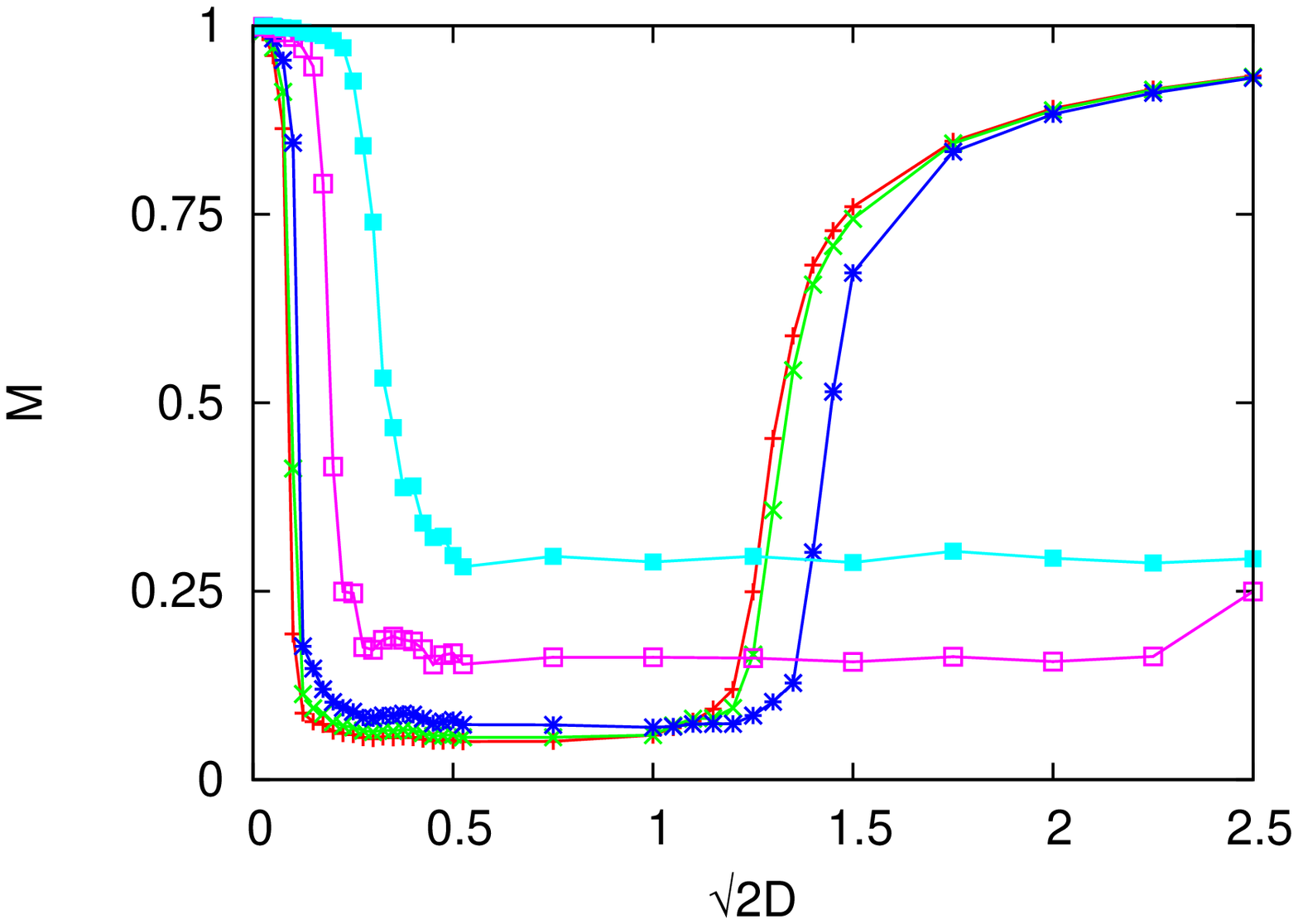}
}
\subfigure[]
{
\label{B}
\includegraphics[width=0.48\textwidth]{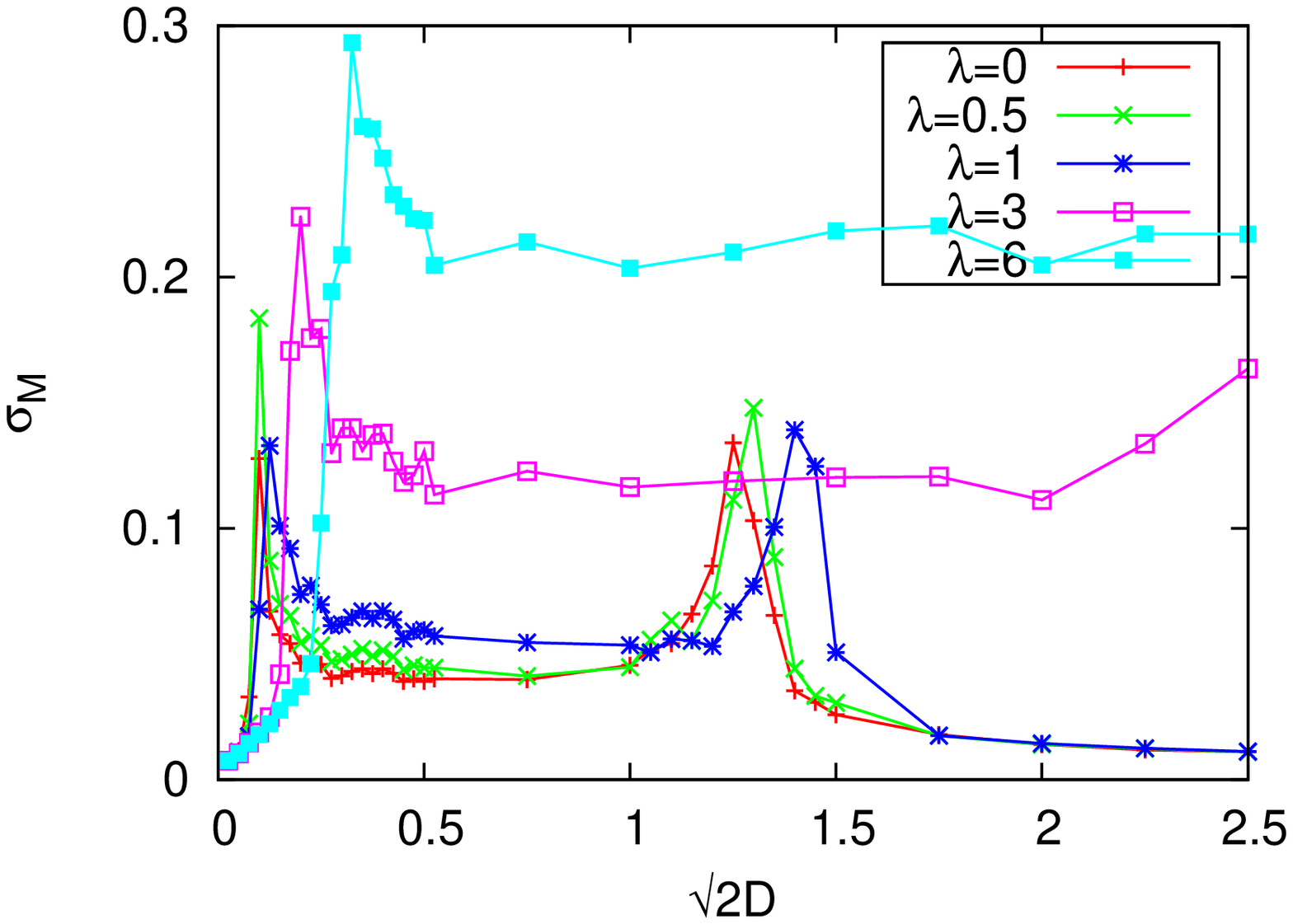}
}
\end{center}
\caption{Re-entrant phase transition in GL model perturbed by additive spatiotemporal sine-Wiener noise for varying white noise strength $\sqrt{2D}$. Initial condition is $\psi(x,0)=1$. Panel \subref{A}: global magnetization $M$. Panel \subref{B}: relative fluctuation $\sigma_M$. Other parameters $B=2.6$ and $\tau=2$.
}
\label{fig_GLeps}
\end{figure}

By varying the white noise strength $\sqrt{2D}$ a re-entrant transition is observed, see fig. \ref{fig_GLeps}.  Note that $\lambda$ increases the lower value of $M$ and shifts the first transition point, whereas its effect on the second transition point - where it exists - is modest.

%\FloatBarrier

\begin{figure}
\begin{center}
\includegraphics[width=0.48\textwidth]{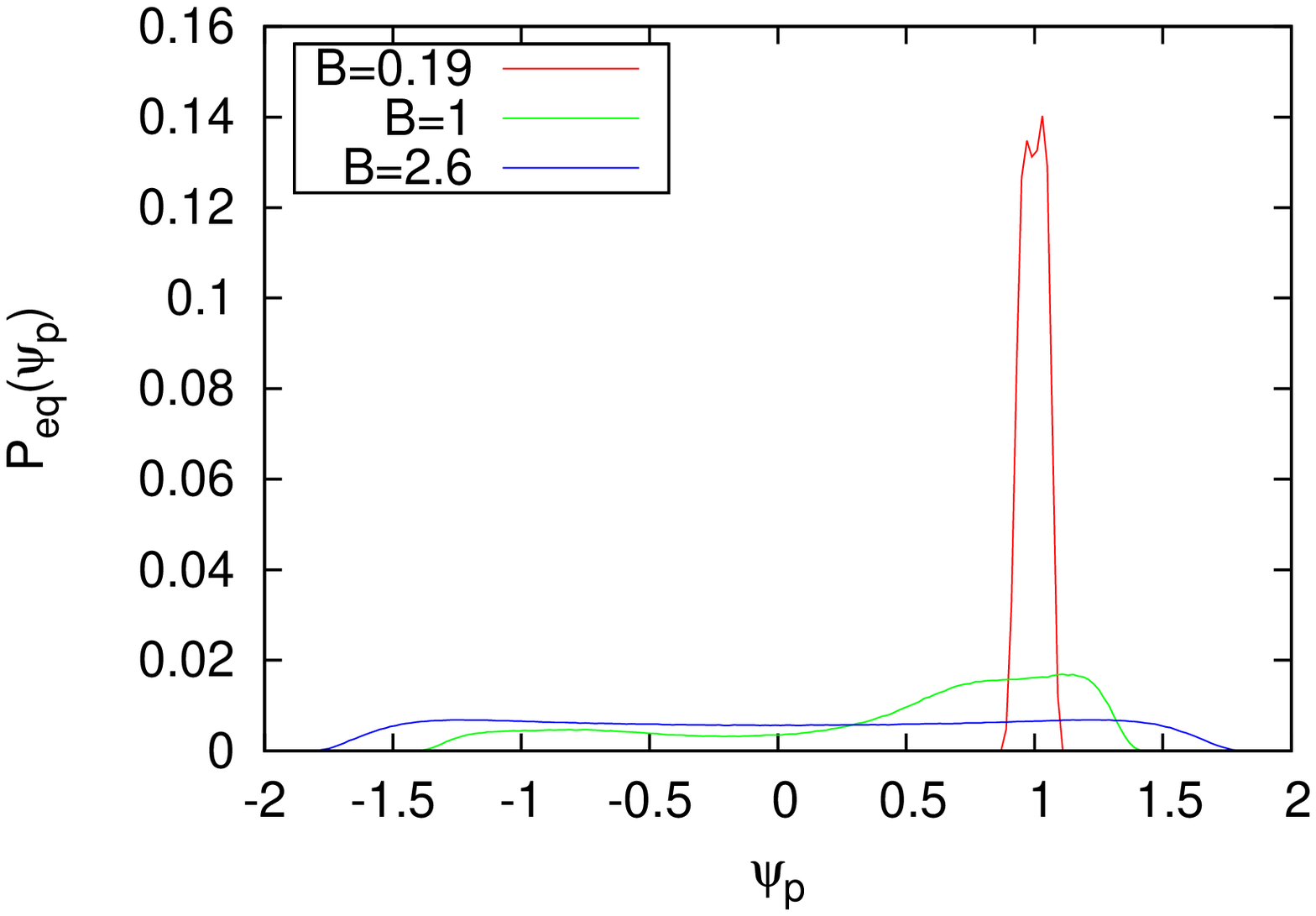}
\includegraphics[width=0.48\textwidth]{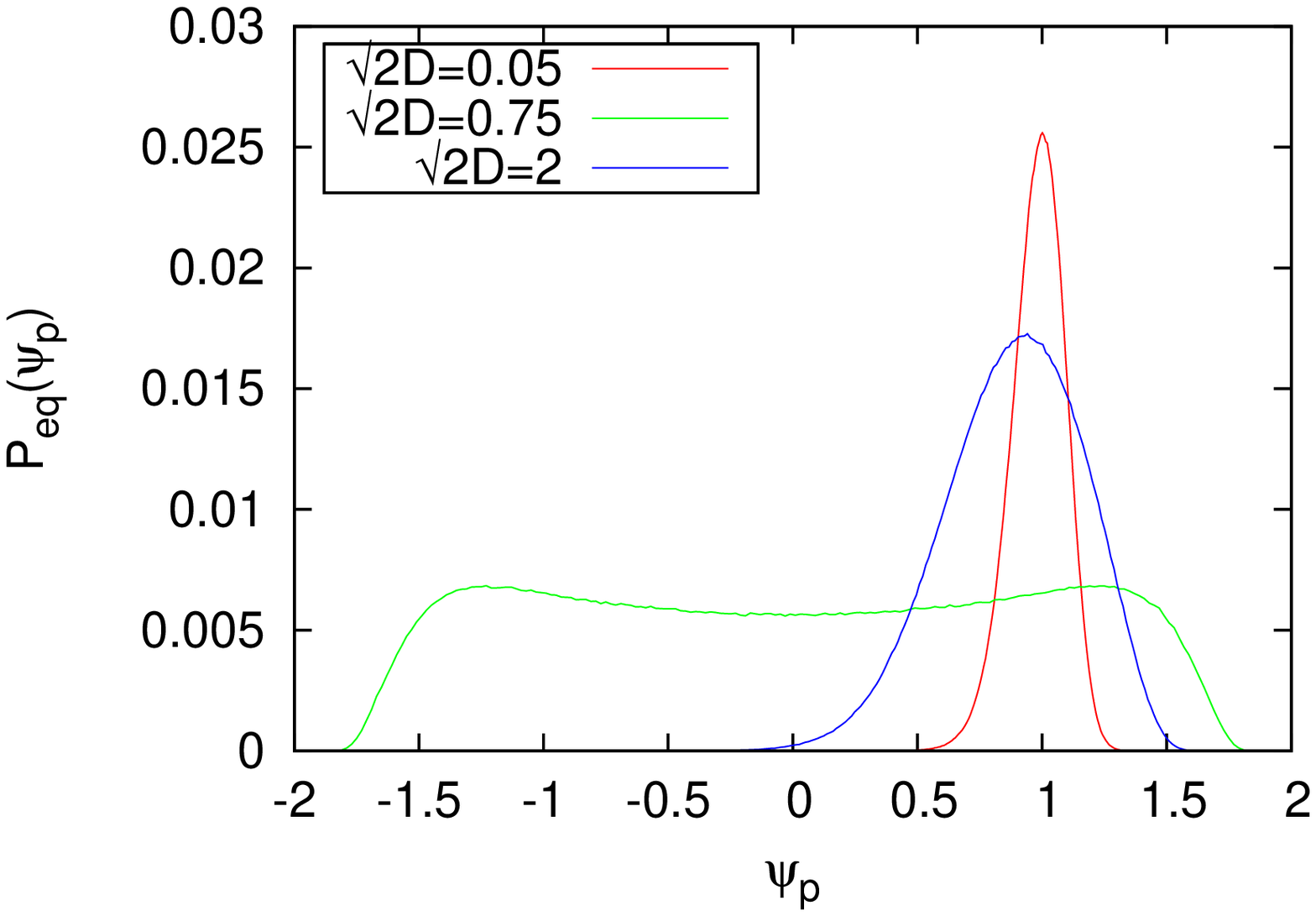}
\end{center}
\caption{Stationary distribution of the field for the GL model perturbed by additive spatiotemporal sine-Wiener noise, in response to changes in noise parameters $B$ (left panel), and $D$ (right panel). Other parameters are, respectively,  ($\tau=2$, $\lambda=1$,  $\sqrt{2D}=0.75$) and ($\tau=2$, $\lambda=1$, $B=2.6$).
}
\label{fig_GLdistri}
\end{figure}

Figure \ref{fig_GLdistri} illustrates the impact of the noise bound $B$ (left panel) and of the white noise strength $D$ (right panel) on the stationary distribution of the field $\phi$ of the GL model. Varying $B$ one may observe transitions from bimodality located close to $\Psi=1$ to bimodality with modes roughly at $\psi =\pm 1.25$. On the contrary,  varying $D$ a re-entrant transition unimodality to bimodality back to unimodality  is observed, which is in line with the re-entrant phase transition showed in fig. \ref{fig_GLeps}. 

%\FloatBarrier

\subsection{Transitory analysis}
\begin{figure}
\begin{center}
\includegraphics[width=8cm]{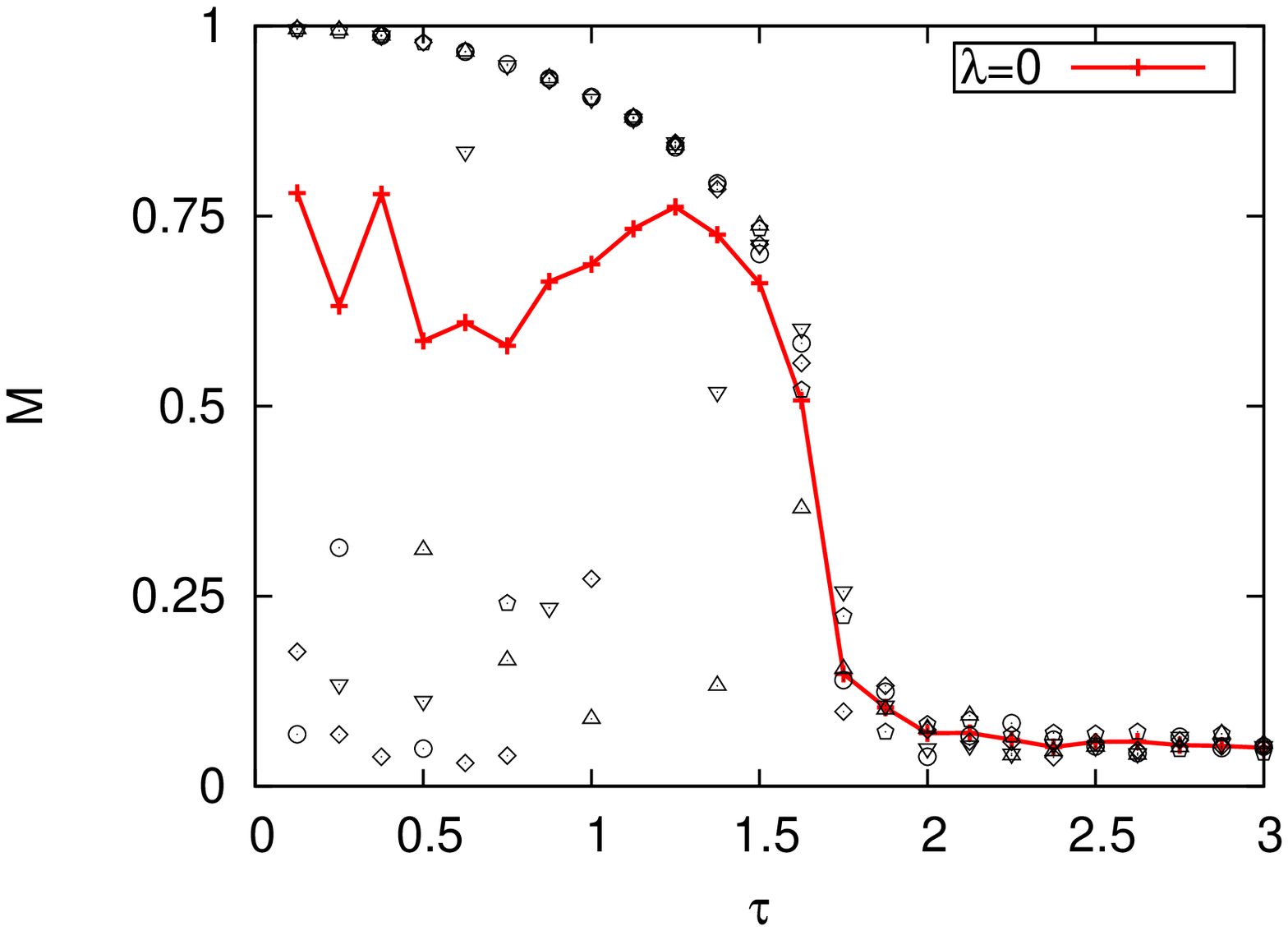}\includegraphics[width=8cm]{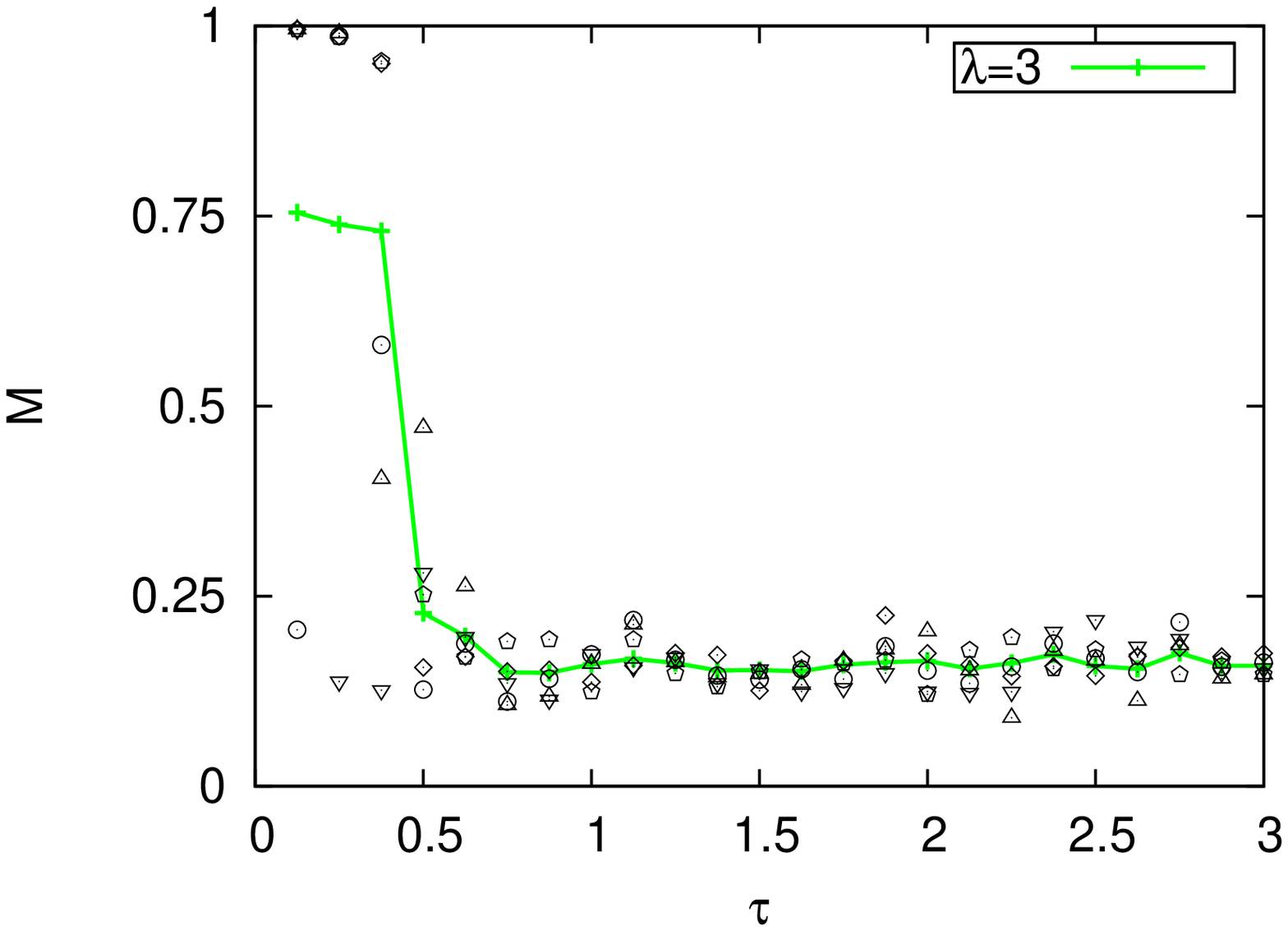}
\end{center}
\caption{Effects of temporal autocorrelation $\tau$ on GL model perturbed by additive spatiotemporal sine-Wiener noise, with disordered initial conditions normally distributed as follows: $\psi_p(0) \propto N(0,0.2)$. Other parameters (as in figure 2): $B=2.4$ and $\sqrt{2D}=1$. 
Dots series represent (with various symbols) different realizations of the system at simulation time $750$, while the continuous line is the average value, computed over all the realizations and over the last 125 time units.}
\label{fig_GLtauIni0}
\end{figure}

In order to study the efficacy of the system in recovering the state with 'large' $M$, we re-consider the same transitions in $\tau$ and $\sqrt{2D}$ formerly analyzed with different initial conditions. Namely, here we assume that at the $\psi_p(0)$ are normally distributed  with zero mean and standard deviation equal to $0.2$: $\psi_p(0) \propto N(0,0.2)$. The simulations and the averages were done, respectively, in the time intervals  $[0,750]$ and $[625,750]$. Figure \ref{fig_GLtauIni0} shows, in the region with large $M$, the onset of very long transient states. On the contrary, the disordered phase is reached after a short transient. Similar results are obtained when varying $D$ (not shown).
%\FloatBarrier

\subsection{Comparison with the GSR noise}
In this section we shall compare the statistical outcomes of the solutions of the GL model under 'equivalent' GSR and sine-Wiener noises, and we shall describe the criteria to establish the related 'equivalence'. 

\begin{figure}
\begin{center}
\includegraphics[width=8cm]{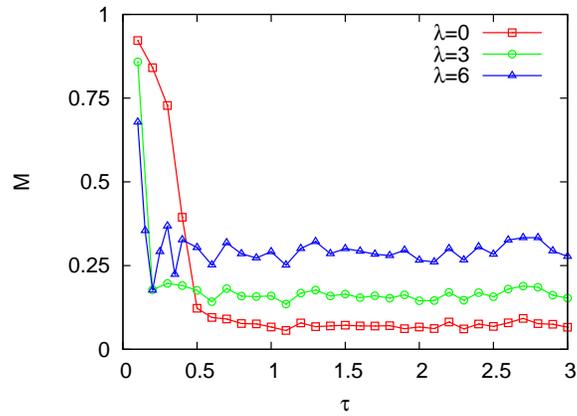}
\end{center}
\caption{Eeffects on GL model of GSR noise obtained by means of formula
(\ref{compD}): curve $M$ vs $\tau$ for GL model perturbed by an additive GSR noise. Other parameters $B=2.4$. To be compared with fig. 2}
\label{F2replypippoz}
\end{figure}

\begin{figure}
\begin{center}
\includegraphics[width=8cm]{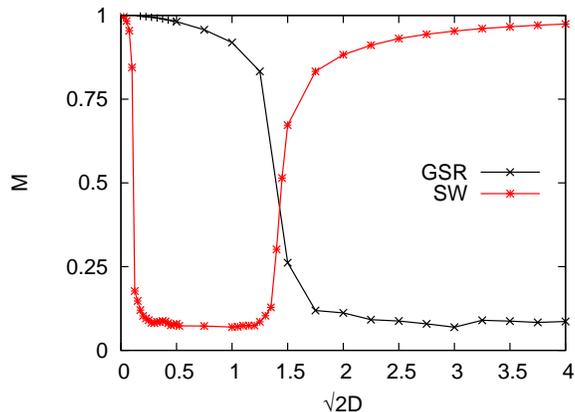}
\end{center}
\caption{
Comparison via eq. (\ref{comptau}) of the effects of GSR vs. sine-Wiener noises on GL model:  $M$ vs $\sqrt{2 D}$. In both curves $\lambda = 1$ and $B=2.6$, and in GSR noise $\tau=2$ . In the GSR noise perturbed case no re-entrant phase transitions are present.}
\label{paperino}
\end{figure}

Let us initially consider fig 2, reporting phase transitions in $\tau$ caused by sine-Wiener noise.  We recall that in fig 2 each curve is identified by a specific value of $\lambda$ and all curves share two parameters: $B= 2.4$ and $\sqrt{2 D} = 1$. How to choose an 'equivalent' GSR noise? The first na\"{i}ve choice would be considering the GSR noise employed as argument of the sinus to generate the sine-Wiener noise. However, this choice would be 'unfair' for the GSR noise, since its span  -roughly quantifiable as the double of its standard deviation $\sigma_{GSR}$ - would be too small. Instead, a more 'fair' way is to adopt GSR noises such that their 'span' is equal to the amplitude of the bounded noise: $2 \sigma_{GSR}=B_{SW}$. As a consequence for the generic $i-th$ point of the $j-th$ curve of fig. 2, identified by the pair $(\tau_i,\lambda_j)$, one has to generate a GSR noise by setting:
\begin{equation}\label{compD} D_{i,j}=\frac{B^2}{4}(1+2 \lambda_j^2)\tau_i. \end{equation}
In other words, at each point we must modify the strength of the white noise that - via the eq. (\ref{XXX}) - generates the sine-Wiener noise. In figure \ref{F2replypippoz} the result of the above-outlined comparison is shown. Also there a phase transition is observed, as the one shown in figure 2, but the transition point is at smaller values of $\tau$.
%\FloatBarrier
Similarly, let us consider the phase transitions illustrated by figure \ref{fig_GLeps}. The k-th point of the j-th curve is defined by the pair $(D_k,\lambda_j)$. Thus, from the relationship $B_{SW}=2 \sigma_{GSR}$ the 'equivalent' GSR noise has to be chosen in a way such that:
\begin{equation} \label{comptau}\tau_{k,j}=\frac{D_k}{(1+2 \lambda_j^2)}\frac{4}{B^2}. \end{equation}
The comparison is shown in figure \ref{paperino}: no re-entrant transition is observed.

%\FloatBarrier

However, a third kind of comparison can be performed by employing a somewhat opposite starting point, which is the response of the GL system to the GSR noise. This type of comparison prescribes that: i) one simulates a GL system excited by a given GSR noise with known parameters, say $(D_x, \tau_x, \lambda_x)$; ii) then one simulates the GL system perturbed by a sine-Wiener noise with the following amplitude
\begin{equation}\label{neweq}
 B_{SW} = 2 \sqrt{ \frac{D_x}{(1+2 \lambda_x^2)\tau_x} }. \end{equation}

\begin{figure}
\begin{center}
\subfigure[]
{
\label{A}
\includegraphics[width=0.48\textwidth]{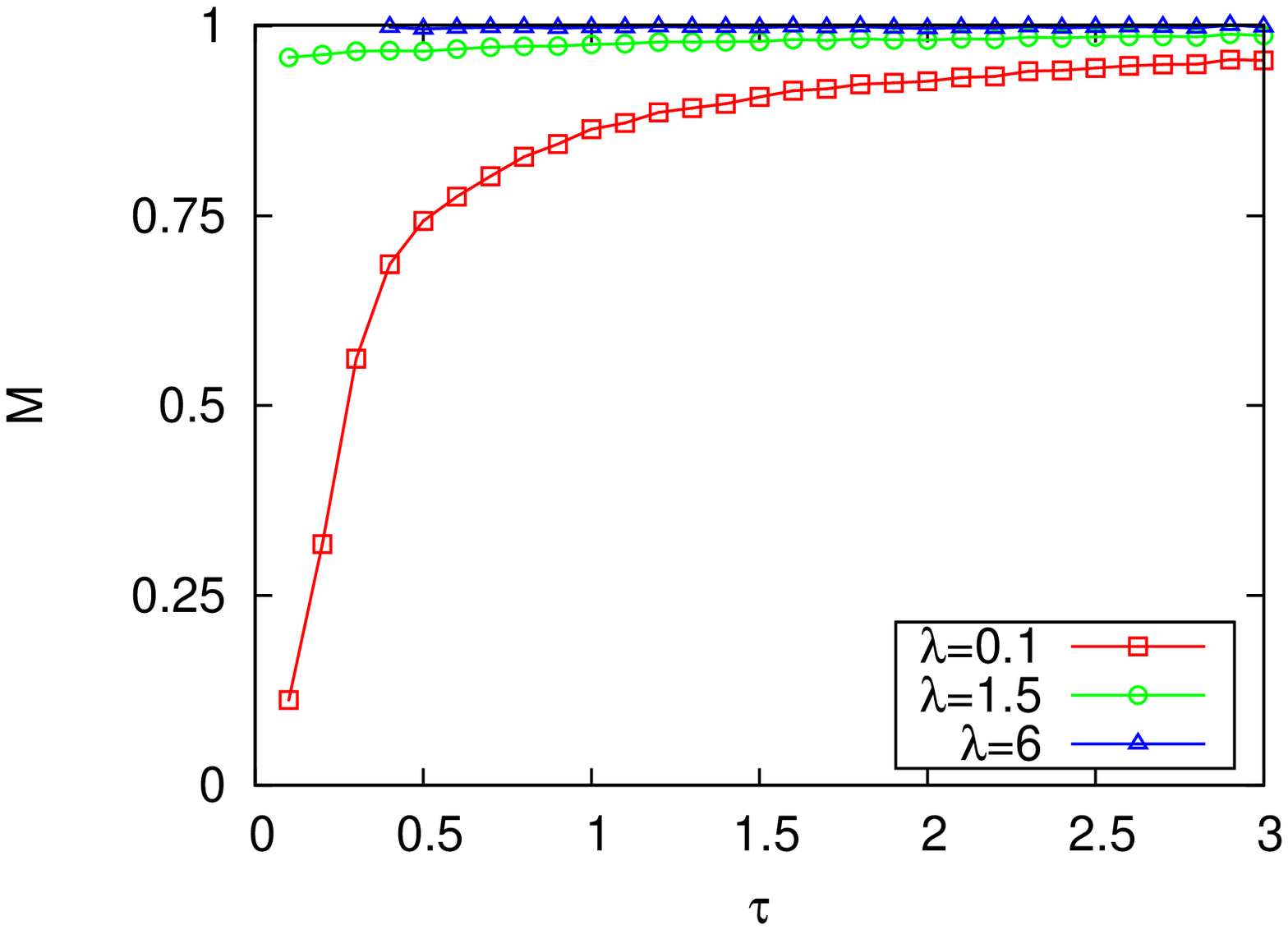}
}
\subfigure[]
{
\label{B}
\includegraphics[width=0.48\textwidth]{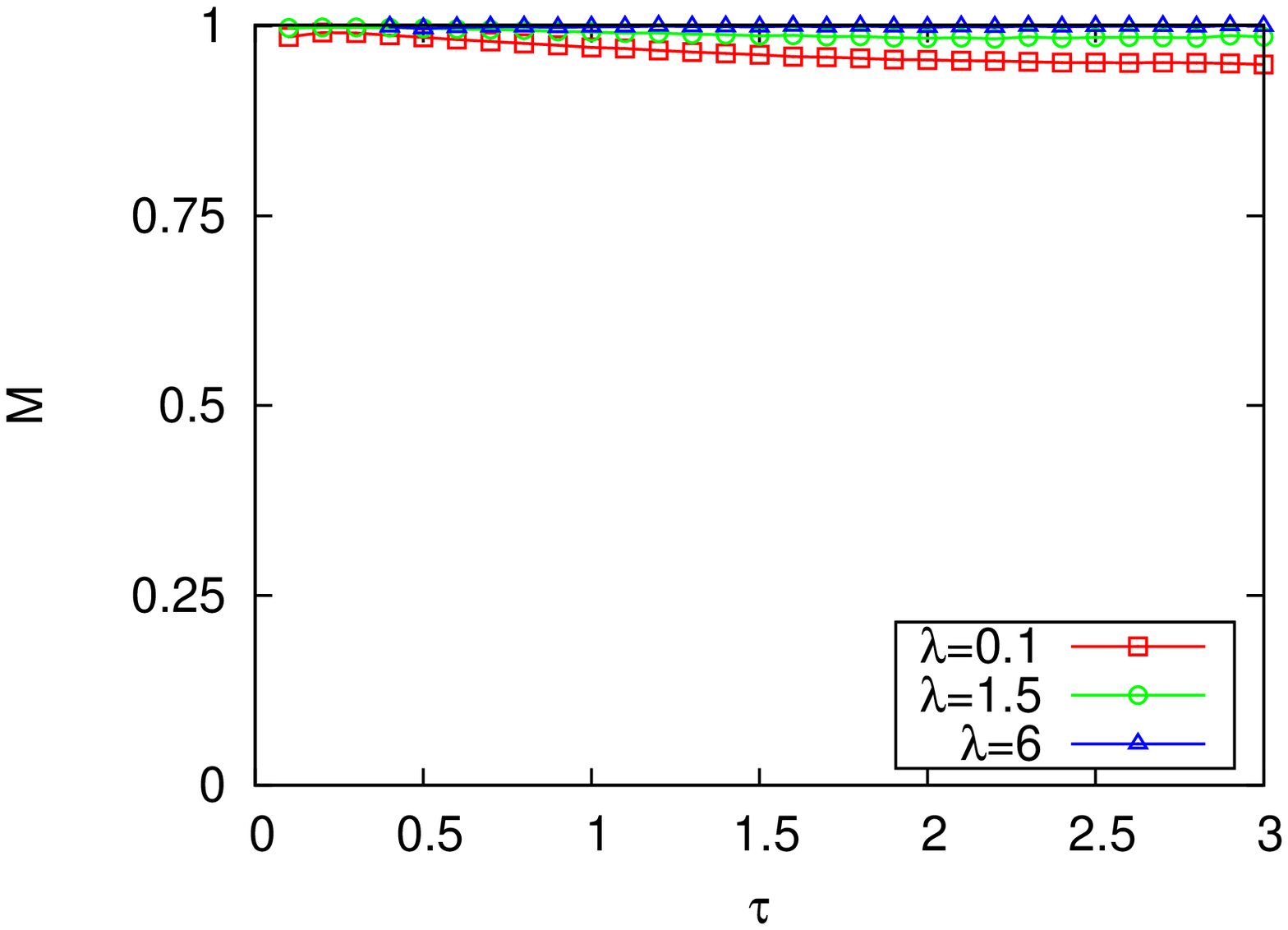}
}
\end{center}
\caption{ Comparison via eq. (\ref{neweq}) of the effects of GSR (panel \subref{A}) vs. sine-Wiener (panel \subref{B}) noises on GL model: $M$ vs $\tau$. For $\lambda =1.5$ and $\lambda =6$ no phase transitions are present in both cases. For $\lambda =0.1$ a transition from low to large values of $M$ can be observed in case of GSR noise, whereas for sine-Wiener noise $M$ remains close to $1$. For GSR noise $\sqrt{2 D}=1$.
}
\label{pluto}
\end{figure}

Figure \ref{pluto}.(a) shows the curve $M$ vs.  $\tau$ for a GL system perturbed by a GSR noise with $\lambda \in \{ 0.1, 1.5,6\}$ and $\sqrt{2 D}=1$. Figure \ref{pluto}.(b) shows the corresponding diagram for the sine-Wiener noise, through the application of formula (\ref{neweq}). As one may see, for $\lambda =0.1$ the GSR noise induces a transition from small to large values of $M$, which is not observed in case of sine-Wiener noise. For larger $\lambda$  both the noises does not induce transitions in $M$. 
%\FloatBarrier

\begin{figure}[t]
\begin{center}
\includegraphics[width=0.48\textwidth]{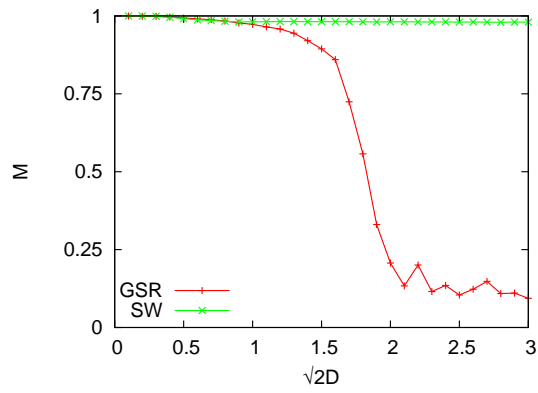}
\end{center}
\caption{Comparison via eq. (\ref{neweq}) of the effects of GSR vs. sine-Wiener noises on GL model:  $M$ vs $\sqrt{2 D}$. In both curves $\lambda = 1$, and in GSR noise $\tau=2$.
}
\label{gigio}
\end{figure}

In figure \ref{gigio} it is shown the curve $M$ vs. $\sqrt{2 D}$ corresponding to a GL system perturbed by a GSR noise with $\lambda = 1$ and $\tau=2$, and the corresponding SW noise (obtained via eq. (\ref{neweq}) ) for $\lambda = 1$. In case of GSR noise a phase transition is observed, whereas for the SW noise no transition is observed.
\FloatBarrier

\section{Concluding remarks}
Here we defined a novel spatiotemporal bounded noise, which is derived from the sine-Wiener temporal bounded noise, and from the spatiotemporal unbounded GSR noise. 

By numerical simulations, and by the properties of the variance of the GSR noise, we showed that the SW noise may undergo a stochastic bifurcation assuming as bifurcation parameter $\lambda$, or $D$. Moreover such bifurcation is also observed in $\tau$, but for large values of this parameter.

In \cite{deFradOnpre} we showed that the Cai-Lin noise also undergoes stochastic bifurcations, which however, are of different nature. Indeed, in that case the bifurcation is from bimodality to unimodality, whereas here the transition is from bimodality to trimodality. Moreover, both for Cai-Lin and for Tsallis-Borland noises no bifurcation is observed with increasing $\tau$, whereas in case of sine-Wiener noise the temporal autocorrelation parameter $\tau$ can induce the bimodality/trimodality transition.

Then we studied the role of the defined noise in the additive perturbation of the GL model. We obtained some effects of interest, among which: i) re-entrant phase transitions in $\sqrt{2D}$; ii) transitions from uni-modality to bi-modality in the distribution of the GL field in correspondence to ordered and disordered phases; iii) disordered phase characterized by clusters of the GL field whose size depends on $\lambda$, and whose permanence depends on $\tau$; iv) different temporal length of transient depending on the assumed initial conditions.

We compared the effect of bounded perturbations on GL systems with those relative to unbounded GSR perturbation with the same fluctuation statistics and spatiotemporal features. This investigation allowed us to stress out, with both numerical simulation and analytical considerations, that the boundedness of noise is crucial for the stability of the 'ordered' state.

The phase transition in $\tau$ observed here share some features with those induced by Cai-Lin spatiotemporal noise \cite{deFradOnpre}, but the transition point occurs in different ranges.

It follows that the observed phenomenologies strongly depend  on the specific model of noise that has been adopted. Then in absence of experimental data on the distribution of the stochastic fluctuations for the problem in study, could be necessary to compare multiple kinds of possible stochastic perturbations models. This is in line with similar observations concerning bounded noise-induced-transitions in zero-dimensional systems \cite{pre}.

Finally, here we faced a systematic comparison between the defined bounded noise and the GSR unbounded noise.

As far as the future investigations are concerned, the priority will be given to a real understanding of the physics underlying the observed phase transitions. 

Moreover, we want to point out the need to perform analytical studies in order to exactly characterize the origin of the bifurcations for the different the sine-Wiener and other bounded noises. This might be important for the above-mentioned physical investigation of the bounded noise-induced phase transitions.

Finally, following the second 'recipe' one might define an entire wide family of spatiotemporal noises derived from the GSR noise, as follows:
$$ \zeta(x,t)= f\left( 2 \ \pi \ \xi(x,t) \right),$$
where $\xi(x,t)$ is a GSR noise, and $f(u)$ is a bounded continuous function. In a further work we shall compare the results here illustrated with those obtained by varying the specific function $f(.)$. 

\section*{Acknowledgments} %We thank very much the five anonymous referees, whose suggestions helped us to greatly improve this work. 
This research was performed under the partial support of the Integrated EU project \textit{P-medicine - From data sharing and integration via VPH models to personalized medicine} (Project No. 270089),
which is partially funded by the European Commission under
the Seventh Framework program.

%%\FloatBarrier
%\section*{APPENDIX}

%\bibliographystyle{ieeetr}
\bibliographystyle{spphys}  
\bibliography{Bibliography} 

\begin{thebibliography}{10}
\providecommand{\url}[1]{{#1}}
\providecommand{\urlprefix}{URL }
\expandafter\ifx\csname urlstyle\endcsname\relax
  \providecommand{\doi}[1]{DOI \discretionary{}{}{}#1}\else
  \providecommand{\doi}{DOI \discretionary{}{}{}\begingroup
  \urlstyle{rm}\Url}\fi

\bibitem{gammaitoni}
L.~Gammaitoni, P.~H\"anggi, P.~Jung, F.~Marchesoni, Rev. Mod. Phys.
  \textbf{70}, 223 (1998)

\bibitem{lucafrancesco}
L.~Ridolfi, P.~D’Odorico, F.~Laio, \emph{Noise-induced Phenomena in the
  Environmental Sciences} (Cambridge University Press, 2011)

\bibitem{hl}
W.~Horsthemke, R.~Lefever, \emph{Noise-Induced Transitions: Theory and
  Applications in Physics, Chemistry, and Biology (Springer Series in
  Synergetics)} (Springer)

\bibitem{wiolindenberg}
H.S. Wio, K.~Lindenberg, Modern Challenges in Statistical Mechanics. AIP
  Conference Proceedings \textbf{658}, 1 (2003)

\bibitem{Ibanhes}
R.T. M.~Iba{\~n}es, J. Garc\'ia-Ojalvo, J.M. Sancho, Lecture Notes in Physics
  \textbf{557/2000}, 247 (2000)

\bibitem{GObook}
J.~Garc\'ia-Ojalvo, J.M. Sancho, \emph{Noise in Spatially Extended Systems}
  (Springer, 1996)

\bibitem{sagues}
F.~Sagu\'es, J.~Sancho, J.~Garc\'ia-Ojalvo, Reviews of Modern Physics
  \textbf{79}(3), 829 (2007)

\bibitem{Wang1}
Q.Y. Wang, Q.S. Lu, G.R. Chen, Physica A: Statistical Mechanics and its
  Applications \textbf{374}(2), 869 (2007)

\bibitem{Wang2}
Q.Y. Wang, Q.S. Lu, G.R. Chen, The European Physical Journal B - Condensed
  Matter and Complex Systems \textbf{54}(2), 255 (2006)

\bibitem{Wang3}
Q.Y. Wang, M.~Perc, Q.S. Lu, S.~Duan, G.R. Chen, International Journal of
  Modern Physics B \textbf{24}, 1201 (2010)

\bibitem{sanchoPhysD}
J.~Sancho, J.~Garc\'ia-Ojalvo, H.~Guo, Physica D: Nonlinear Phenomena
  \textbf{113}(2-4), 331  (1998)

\bibitem{hanggi}
P.~Jung, P.~H\"anggi, Phys. Rev. A \textbf{35}, 4464 (1987)

\bibitem{GO92}
J.~Garc\'ia-Ojalvo, J.M. Sancho, L.~Ram\'irez-Piscina, Phys. Rev. A
  \textbf{46}, 4670 (1992)

\bibitem{lam}
P.M. Lam, D.~Bagayoko, Phys. Rev. E \textbf{48}, 3267 (1993)

\bibitem{GO94}
J.~Garc\'ia-Ojalvo, J.M. Sancho, Phys. Rev. E \textbf{49}, 2769 (1994)

\bibitem{New05}
M.E.J. Newman, Contemporary Physics \textbf{46}(5), 323 (2005)

\bibitem{wioII}
H.S. Wio, R.~Toral, Physica D: Nonlinear Phenomena \textbf{193}(1-4), 161
  (2004)

\bibitem{CaiLin}
G.Q. Cai, Y.K. Lin, Phys. Rev. E \textbf{54}, 299 (1996)

\bibitem{bobryk}
R.V. Bobryk, A.~Chrzeszczyk, Physica A: Statistical Mechanics and its
  Applications \textbf{358}(2-4), 263  (2005)

\bibitem{dimentberg}
M.~Dimentberg, \emph{Statistical dynamics of nonlinear and time-varying
  systems} (Research Studies Press, 1988)

\bibitem{Homburg}
H.~Zmarrou, A.J. Homburg, Ergodic Theory and Dynamical Systems \textbf{27}(05),
  1651 (2005)

\bibitem{pre}
A.~d'Onofrio, Phys. Rev. E \textbf{81}, 021923 (2010)

\bibitem{dongan}
A.~d'Onofrio, A.~Gandolfi, Phys. Rev. E \textbf{82}, 061901 (2010)

\bibitem{deFradOnpre}
S.~de~Franciscis, A.~d'Onofrio, Phys. Rev. E \textbf{86}, 021118 (2012)

\bibitem{GO92pla}
J.~Garc\'ia-Ojalvo, J.~Sancho, L.~Ram\'irez-Piscina, Physics Letters A
  \textbf{168}(1), 35  (1992)

\bibitem{gpsv}
J.~Garc\'ia-Ojalvo, J.M.R. Parrondo, J.M. Sancho, C.~Van~den Broeck, Phys. Rev.
  E \textbf{54}, 6918 (1996)

\bibitem{various}
O.~Carrillo, M.~Iba\~nes, J.~Garc\'ia-Ojalvo, J.~Casademunt, J.M. Sancho, Phys.
  Rev. E \textbf{67}, 046110 (2003)

\bibitem{ms}
R.S. Maier, D.L. Stein, Proc. SPIE-The Int. Soc. for Optical Engineering
  \textbf{5114}, 67  (2003)

\bibitem{jstat}
N.~Komin, L.~Lacasa, R.~Toral, Journal of Statistical Mechanics: Theory and
  Experiment \textbf{2010}(12), P12008

\bibitem{ss}
S.~Scarsoglio, F.~Laio, P.~D'Odorico, L.~Ridolfi, Mathematical Biosciences
  \textbf{229}(2), 174  (2011)

\bibitem{ouch}
K.~Ouchi, N.~Tsukamoto, T.~Horita, H.~Fujisaka, Phys. Rev. E \textbf{76},
  041129 (2007)

\bibitem{parrondo}
C.~Van Den~Broeck, J.M.R. Parrondo, R.~Toral, Physical Review Letters
  \textbf{73}(25), 3395 (1994)

\bibitem{coppel}
W.~Coppel, \emph{Asymptotic Behavior of Differential Equations} (Heath, 1965)

\bibitem{rt}
N.~Komin, L.~Lacasa, R.~Toral, Journal of Statistical Mechanics: Theory and
  Experiments \textbf{10}, 12008 (2010)

\end{thebibliography}
\end{document}